\documentclass[prd,aps,showpacs,preprint]{revtex4}
\usepackage{dcolumn}
\usepackage{graphics}
\newcommand{\feynslash}[1]{#1\hspace{-10pt} \slash \hspace{6pt}}

\makeatletter

\let\SF@@footnote\footnote
\def\footnote{\ifx\protect\@typeset@protect
    \expandafter\SF@@footnote
  \else
    \expandafter\SF@gobble@opt
  \fi
}
\expandafter\def\csname SF@gobble@opt \endcsname{\@ifnextchar[
  \SF@gobble@twobracket
  \@gobble
}
\edef\SF@gobble@opt{\noexpand\protect
  \expandafter\noexpand\csname SF@gobble@opt \endcsname}
\def\SF@gobble@twobracket[#1]#2{}
\newcommand{\eqr}[1]{Eq.~(\ref{#1})}
\newcommand{\tbarsq}{\bar{T}_f^2 +\bar{T}_{f_2}^2}
\newcommand{\vscale}{\left(\frac{v}{180\,{\rm GeV}}\right)}

\makeatother

\begin{document}
\input epsf

\title{
Standard Model baryogenesis through four-fermion operators in braneworlds}

\author{Daniel J.\ H.\ Chung}\thanks{email:daniel.chung@cern.ch}
        \affiliation{CERN Theory Division, \\CH-1211 Geneva 23, Switzerland}
\author{Thomas Dent}\thanks{email:tdent@umich.edu}
        \affiliation{Michigan Center for Theoretical Physics,\\
University of Michigan, Ann Arbor, MI 48109}

\date{December 2001}

\begin{abstract}\noindent
We study a new baryogenesis scenario in a class of braneworld models
with low fundamental scale, which typically have difficulty with 
baryogenesis. The scenario is characterized by its minimal nature: the 
field content is that of the Standard Model and all interactions
consistent with the gauge symmetry are admitted. 
Baryon number is violated via a dimension-6 proton decay operator, 
suppressed today by the mechanism of quark-lepton separation in extra 
dimensions; we assume that this operator was unsuppressed in the 
early Universe due to a time-dependent quark-lepton separation. The 
source of $CP$ violation is the CKM matrix, in combination with the 
dimension-6 operators. We find that almost independently of cosmology, 
sufficient baryogenesis is nearly impossible in such a scenario if the 
fundamental scale is above 100 TeV, as required by an unsuppressed
neutron-antineutron oscillation operator. The only exception
producing sufficient baryon asymmetry is a scenario involving 
out-of-equilibrium $c$ quarks interacting with equilibrium $b$ quarks.
\end{abstract}
\maketitle
\section{Introduction}
The experimental lower limit on the lifetime of the proton is a severe
problem for models in which the fundamental scale of quantum gravity
is low compared to the supersymmetric GUT scale $10^{16}\,$GeV
\cite{Lykken,AHDD+IAHDD}.  A baryon number U$(1)$ symmetry cannot be
gauged in field theory; like other accidental symmetries it is
expected to be violated by effects at the string scale or by quantum
wormholes and virtual black holes \cite{BanksDixon_Gilbert}. Such $B$
violation would appear at low energies as nonrenormalizable operators,
for example of the form $\lambda qqql/M_f^2$ where $M_f$ is the
fundamental scale and $\lambda$ is expected to be ${\mathcal O}(1)$ in
the absence of suppression mechanisms. Then for $M_f$ in the
${\mathcal O}(10-100)\,$TeV range, for which collider signals of the
fundamental degrees of freedom or of large extra dimensions may be
observable, $\tau_p$ comes out to be under a second, to be compared
with the (mode-dependent) experimental bounds of order $10^{32}$ {\em
years} \cite{Groom:in}. Various solutions have been proposed 
\cite{benakli,ahs,Ibanez}
all of which have implications for the production of an excess of
baryons over antibaryons in the early Universe, for which $B$
violation is a precondition.

The production of gravitational Kaluza-Klein modes in the early Universe 
in such models, in which some compactified extra dimensions are orders of 
magnitude larger than the fundamental length, gives a severe upper bound on 
the maximum temperature attained consistent with cosmological observations. 
Even for the maximum number (usually considered as 6) of large extra 
dimensions and the relatively large value $M_f = 100\,$TeV, a maximum
temperature of a few GeV is the upper bound if overclosure of the Universe, 
disruption of the successful predictions of nucleosynthesis, and an 
observationally unacceptable level of background gamma-rays from K-K mode 
decay are to be avoided \cite{benakli,Hall+malc+hannestadcosmo}. The energy
density in K-K modes would also affect the evolution of density fluctuations 
by altering the time of matter-radiation equality \cite{Fairbairn:2001ab}. 
Astrophysical production and decay of such modes also leads to an independent 
lower bound on the fundamental scale \cite{CullenBarger_nHannestad}, which is 
also constrained by the non-observation (to date) of direct and loop effects 
in experiments \cite{KKsigs}. 

Any attempt at explaining proton longevity and baryogenesis should operate 
within these constraints. Exact (anomaly-free) discrete or horizontal 
symmetries can be imposed to forbid $B$-violating operators mediating proton 
decay \cite{KraussDGS,benakli} while allowing others, through which 
baryogenesis occurs: this approach requires an ``X-boson'' to be present, 
with couplings which appear unnaturally small (in contrast to the standard 
GUT or leptogenesis scenarios). Baryon number can be gauged if the anomaly 
is canceled by a string theory mechanism, or $B$ violation may be forbidden 
to all orders in perturbation theory by string selection rules, in some 
``intersecting brane'' models \cite{Ibanez}. Note however that in a 
more general class of intersecting brane models \cite{kors_etal}, such 
selection rules do not prevent the four-fermion operators from appearing, 
as discussed above, in which case the fundamental scale cannot be low.

If baryon number is perturbatively exact, nonperturbative processes 
\cite{DavidsonLR} are the only option to create net baryon number.
It is difficult to see how this proposal can be 
reconciled with cosmological constraints, since any such processes would 
operate at or above the electroweak scale and be enormously suppressed at 
low temperature \footnote{It is unclear whether electroweak baryogenesis 
with low $T_{\rm rh}$ \cite[first reference]{DavidsonLR} actually 
satisfies the bounds, since a hot plasma at temperature $T\gg T_{\rm rh}$
is needed.}.

A novel mechanism for suppressing proton decay in the context of a low
fundamental scale theory has been proposed in Ref.~\cite{ahs}.  This
is a geometrical mechanism for suppressing 4D $B$-violating operators,
namely localization of fermions in extra dimensions.  The simplest
implementation is for the SU$(2)_L\times {\rm U}(1)_Y$ gauge fields to
propagate in one extra dimension ({\it cf.}\/\ \cite{DienesDG}), in
which the quark and lepton wavefunctions are peaked about points
separated by a distance $L\sim 30 M_f^{-1}$
(Fig.~\ref{fig:branesetup}). Then any strong $B$-violating operators
in the effective 5D theory can only produce proton decay proportional
to the overlap of the wavefunctions, which can be exponentially
small. Alternatively, proton decay by exchange of massive modes is
suppressed by the Yukawa propagator over the distance $L$
\footnote{Assuming that no light fermions with $B$-violating
interactions propagate over the bulk.}. Nonperturbative quantum
effects which may lead to proton decay, for example virtual black
holes \cite{Adams}, are also exponentially suppressed due to the
integration over the fifth dimension.

Baryogenesis in such scenarios has been discussed in Ref.~\cite{Masiero}.  
However, the authors of that work introduced a Standard Model singlet 
scalar with renormalizable baryon number-violating interactions and 
primarily considered the effect of thermal corrections to the wave function 
overlap which controls the strength of the baryon number violation.

An Affleck-Dine-type scenario consistent with large extra dimensions,
which involves introducing a scalar charged under a new $U(1)$ group, has 
recently been described in \cite{Allahverdi:2001dm}; while the idea 
appears viable from a cosmological viewpoint, the authors do not discuss
how $B$-violating couplings which induce proton decay are to be 
suppressed, while still keeping the operators by which the scalar decays. 
One might imagine that additional exact symmetries could be imposed to 
achieve this, in the spirit of \cite{benakli}.

In this work, we consider a related, but simpler class of scenarios
and report mostly on a ``no go'' result.  Specifically, we consider
the feasibility of baryogenesis under the following conditions:

\begin{enumerate}
\item The only renormalizable operators that are present at the
electroweak scale are those of the Standard Model, and the perturbative 
4D effective field theory (dimensionally reduced from an intrinsically 
higher-dimensional theory) is valid. (The possibility of doing without 
an ``X-boson'' was briefly mentioned in \cite{Masiero}.) In other words, 
we work with an effective
theory with top as the heaviest field that has not been integrated out
and neglect any Lorentz violating operators resulting from time
dependence of the extra dimensional geometry.  This condition is
motivated from our ignorance about the physics beyond the Standard
Model and the fact that no viable baryogenesis scenario has yet been
proposed with only the Standard Model fields and gravity.
\item The leading baryon number-violating operator due to physics at 
the fundamental scale is a dimension-6 operator of the form 
\begin{equation} \label{eq:wavefunctionoverlap}
\frac{v_{s}}{M_{f}^{2}}qqql
\end{equation} 
where \( v_{s} \) the wave function overlap, which is a time dependent 
function. The wave function overlap \( v_{s} \) can be time-dependent 
whenever fermion localization is controlled by a field, akin to the radion, 
which is initially displaced from the minimum of its effective potential. 
The work of \cite{Masiero} can also be taken to be of this type, where the 
``radion'' is always at the minimum of a time dependent effective potential, 
which is derived from time dependent finite temperature effects. 
Note that this condition implies the assumption that 4D 
instanton/sphaleron induced baryon number violating operators are weaker 
than the dimension-6 operator, {\em i.e.}\/\ only perturbative physics
is analyzed for dimension \( 4 \) operators. This is a reasonable 
assumption as long as the temperature is well below the electroweak scale
(which is anyway required in low fundamental scale models) and the 
SU$(2)_L$ coupling remains small (Appendix \ref{app:inst}).
\item The 4D Planck scale \( M_{pl} \) has an adiabatic time
dependence due to the change in the volume of extra dimensions, and
the fundamental scale of gravitational physics is at $M_f \simeq 10^5\,$GeV.
The upper bound on this scale comes from the desire to ameliorate
the hierarchy problem without SUSY, while the lower bound
is obtained by the bounds on neutron-antineutron oscillations, which cannot
be suppressed by a geometrical mechanism. Note that there is still a certain
amount of fine tuning, of order $10^3$, with such a (relatively) large 
fundamental scale.
\item There is no electroweak phase transition because of a low
reheating temperature ({\em i.e.}\/\ $T<30\,$GeV) 
\cite{benakli,Hall+malc+hannestadcosmo}. Baryogenesis
is also assumed to occur above the QCD phase transition temperature
({\em i.e.}\/\ $T>0.2\,$GeV).
\item The effective 3D space is homogeneous and isotropic at all times
when the temperature of the universe is $T>0.2\,$GeV.
\end{enumerate}
In this context, we analyze the possibility of baryogenesis and find
specific conditions under which it is feasible.  

Our setup contains all the necessary ingredients of baryogenesis 
\cite{sakharov}:
\begin{enumerate}
\item Baryon number violation (dimension-6 operator)
\item \( C \) and \( CP \) violation (Standard Model CKM + dimension-6 
operator)
- the resulting quantity is much larger than the Jarlskog invariant.
\item Nonequilibrium (spacetime expansion or other unspecified means)
\end{enumerate}
The reason for the difficulty of this scenario will be ingredient 1,
the baryon number violating operator being too weak in the low
temperature setting, independently of ingredients 2 and 3. Since the 
usual Jarlskog invariant of the SM is not the relevant quantity, given the
presence of dimension-6 operators, the reason why this baryogenesis 
scenario is difficult has nothing to do with traditional wisdom. 
In particular, we find that dimension-6 operators are in most
cases too weak to compete with baryon number-conserving operators
which create entropy, for most sets of reactants.

Na\"\i vely, one finds this result a little surprising because one would expect
\begin{equation}
\eta_{B} \equiv \frac{n_{B}}{s}\sim \frac{\delta_{CPV} Bn}
{g_{*S}T^{3}_{*}}
\end{equation} 
can be easily engineered to be
sufficiently large, where \( \delta_{CPV} \) is the dimensionless
quantity associated with $CP$ violation, \( B \) is the maximal
branching ratio into the baryon number violating channel fixed by our
model, \( n \) is the number density of particles that can be
converted into baryon number, \( g_{*S} \) is the number of degrees of
freedom contributing to the entropy, and \( T_{*} \) is the
temperature at some fiducial time after which the baryon number is
conserved. However, even though \( \delta_{CPV} \) can be naturally
large in our scenario (not governed by the Jarlskog invariant), \(B
n/(g_{*S}T_{*}^{3}) \) cannot in general be made large because as the
baryon number violating reactions are creating baryon asymmetry, the
baryon number conserving counterparts are creating proportionately
large amounts of entropy density.  In other words, having the
quark-lepton separation go to zero does not generate order 1
branching of the baryon number violating channels.  Instead, such
channel is always suppressed by a scale much larger than the
electroweak scale.

We do find an exception to our ``no go'' rule for a special situation
in which the universe starts out dominated by the $b$ and $c$ quarks
and gluons (which might happen through the peculiarities of 
reheating after inflation),
and the lighter quark $c$ is out of equilibrium while the heavier
quark $b$ is in equilibrium, for example due to the anomalously large
expansion rate and the anomalously small density of $c$ quarks.  In
this exceptional scenario, there are two key ingredients compensating
the suppression of the dimension 6 operator:
1. Due to the small $|V_{cb}|^2$, the $b$ quark interactions with out 
of equilibrium $c$ quarks do not generate significant entropy; 2. The 
integration time for baryogenesis can be
made long if the expansion rate is small, due to the nonstandard
cosmology of extra dimensions.  However, this remote possibility,
which requires additional assumptions about special initial conditions
and/or more new physics, is unlikely to be realizable, and we do not
have a detailed model to demonstrate its full viability.

Throughout this paper, we do not assume thermal equilibrium initial 
conditions, except where explicitly noted.
Hence, although the short distance operators are fixed, the nonzero
initial particle densities that exist in the universe can be {\em a priori}\/
arbitrary combinations of species.  Thus, when we consider bounds, we
explicitly check all likely combinations of species without any bias
to their initial number distributions, except when otherwise noted. As
far as the out-of-equilibrium-condition due to the expansion of the
universe is concerned, we assume that somehow the Planck scale can be
adjusted to dial into the appropriate expansion rate.  Of course, in a
more completely specified model, this may turn out to be problematic, due
to the bounds on moduli overproduction and other light moduli problems.  
However, our results are robust with respect to how the moduli
problems and initial conditions problems can be solved.  In fact most
of our results do not depend on how the particles are out of
equilibrium.  This broad insensitivity with respect to cosmological
details is one of the key features to our work.

The order of our presentation will be as follows.  We begin with a
discussion of the class of braneworlds and the energy scales and 
couplings that we are interested in.  We then present our constraint on 
this scenario for baryogenesis with dimension-6 operators in the low 
temperature regime. Finally, we summarize and conclude.

\section{Braneworld setup and scales}
The simplest way to realize the proposal of geometrical proton decay 
suppression is for the ``brane'' on which we live to
be a nontrivial scalar field profile in an extra dimension \cite{Rubakov:bb}.
On a smooth manifold, chiral fermions are localized in the presence of a 
scalar field profile with Yukawa couplings to the fermion fields, resulting 
in a position-dependent effective mass term. The simplest ansatz is to take 
a linearly varying scalar field, which results in chiral fermion zero modes 
with a Gaussian wavefunction, localized about the point where the effective 
fermion mass vanishes \cite{ahs}. On an orbifold \cite{ggh} one may impose 
special boundary conditions, which themselves ensure that the lowest mode is 
chiral, then with a scalar field which is forced to take up a kink 
configuration one obtains a fermion wavefunction which varies approximately 
as $\cosh^{-2}(kx_5)$. 

By coupling the leptons and quarks differently to different scalar fields, 
one may separate the zero mode wave functions such that the wave
function overlap factor $v_s$ in \eqr{eq:wavefunctionoverlap} obtains
an exponentially suppressed value.
As discussed in \cite{ggh}, it is not easy to induce {\em large}
displacements of the points of localization away from special points
such as the orbifold fixed points (whose boundary conditions force the
zero of the scalar field) without introducing arbitrary scale parameters.
In this paper, we will not assume any particular mechanism to
make these localization points (which we will sometimes loosely call
``branes'') dynamical, but rather suppose that large dynamical
displacements of the localization points (branes) from their final
value are possible.  Hence, we treat $v_s$ as merely a time-dependent 
function that is controlled by an exponential of the form
\begin{equation}
\label{eq:defzeta}
v_s \sim e^{-\zeta M_f L}
\end{equation}
where $\zeta$ is a constant coefficient of order 1 and $L$ is a 
time-dependent function giving the distance between the quark and lepton 
localizations. There may be additional contributions to $v_s$ varying as
$e^{-(\mu L)^2}$ but these will be subleading unless $\mu^2\ll M_f^2$.
We will see below that $\zeta$ actually has to be greater than or equal to 
about $3/2$ in our scenario. 

The baryogenesis scenario is then as follows: at some early time in the
evolution of the Universe, $M_f L\sim {\mathcal O}(1)$ and the baryon 
number violating operator is suppressed only by the fundamental scale.  
By the time of nucleosynthesis, $\zeta M_f L$ will have
settled down to today's value of about 50, consistent with proton decay
constraints (see Appendix \ref{app:pprotection}).  
For the purposes of our paper, we will not need the specific details
of how the quark-lepton separation is stabilized at its final value.
We will show that the short-distance physics prevents sufficient
baryogenesis in the minimal model that we consider, except in the special
case already mentioned. 

Since the zero mode quark and lepton wave functions are separated in
the extra dimension and since the quarks and the leptons interact
through the $SU(2)_L\times U(1)_Y$ gauge group, these gauge fields
must also propagate in the extra dimension, and the 4D gauge theory
must come from dimensional reduction.  Similarly, since everything
interacts with gravity, 4D gravity also arises from dimensional
reduction.  We now derive bounds on the fundamental scale as well as
the size of the extra dimensions by demanding perturbativity of gauge
fields propagating in the extra dimensions and that Einstein gravity
be recovered \footnote{We also assume that the brane tension is
negligible such that Randall-Sundrum type of warping is negligible.}.
 
\begin{figure}[t]
{\par\centering \resizebox*{0.9\columnwidth}{!}
{\includegraphics{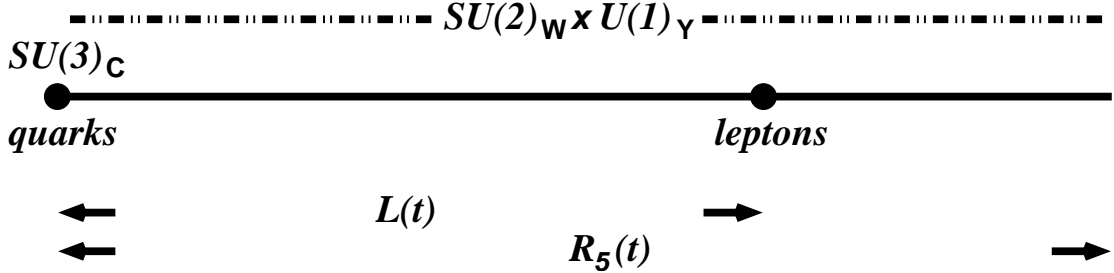}} \par}
\caption{\label{fig:branesetup} Schematic setup of our braneworld.}
\end{figure}

From gravitational dimensional reduction, given that there are a total
of \( n \) extra dimensions and one of them is asymmetric with length
\( R_5(t_3) \), the others having lengths $R_i$, $i=1,\ldots,n-1$, we have 
\begin{equation} \label{gravityscale}
M_{f}^{2}=\frac{\bar{M}_{pl}^{2}}{(R_5(t_3)M_{f})\prod _{i}^{n-1}
(R_{i}M_{f})}
\end{equation}
where $t_3$ denotes the present time and $\bar{M}_{pl}$ is the reduced
Planck mass $M_{pl}/\sqrt{8\pi}$.
There is also a perturbativity constraint on the Yang-Mills coupling constant.
In \( 4+p \) dimensions (for example for a \( (3+p) \)-brane), the Yang-Mills
coupling is 
\[ g_{4+p}^{2}=\frac{\lambda ^{2}_{4+p}}{M_{f}^{p}}\]
where \( \lambda _{4+p} \) is a dimensionless number. By dimensional reduction
to 4D, we have the 4-dimensional dimensionless Yang-Mills coupling \( g_{4} \)
related to \( \lambda _{4+p} \) by 
\begin{equation} \label{eq:4dandhigher}
\lambda _{4+p}^{2}=g^{2}_{4}(R_5M_{f})\prod _{i}^{p-1}(R_{i}M_{f}).
\end{equation}
If \( g_{4}(t) \) is to remain perturbative at time \( t_{1}\leq t_3 \), 
we must have
\[ \left( \frac{g^{2}_{4}(t_{3})}{4\pi }\right) \left( \prod _{i}^{p-1}\left(\frac{R_{i}(t_{3})}{R_{i}(t_{1})}\right)\right) \left( \frac{R_5(t_{3})}{R_5(t_{1})}\right) <1. \]
Given that \( g^{2}_{4}(t_{3})/(4\pi )\approx 1/31 \) for the EW theory [\( \alpha \equiv e^{2}/4\pi \simeq 1/137 = \alpha_{W}\sin^{2}\theta_{W} \simeq 0.225\alpha_{W}] \),
we require
\[ R_5(t_{3})/R_5(t_{1})\left( \prod_{i}^{p-1}
\left(\frac{R_{i}(t_{3})}{R_{i}(t_{1})}\right)\right) <31.\]
If we have a scenario in which 
\[ R_5(t)=\textrm{const. }\]
this is not an issue, but in general, this is constraining. 

Now, let us consider absolute perturbativity of the five dimensional theory.
Since the cutoff is set at $M_f$, perturbativity requires
\[ \frac{g_{4+p}^{2}}{4\pi }M_{f}^{p}<1\]
Using Eq.~(\ref{eq:4dandhigher}), we obtain
\begin{equation} \label{5dpertbound}
\frac{\lambda _{4+p}^{2}}{4\pi }=\frac{g_{4}^{2}(t_{3})}{4\pi }(R_5(t_{3})M_{f})\prod _{i}^{p-1}(R_{i}(t_{3})M_{f})<1
\end{equation}
for 5D perturbativity. However, this requires the volume enhancement factor
to be small. In the end, if we want to minimize this effect, we must have 
only a D4-brane ({\em i.e.}\/\ \( p=1 \)). 
Note that our perturbativity constraint is for calculational feasibility 
rather than necessarily for fundamental physical reasons.

Coming back to Eq. (\ref{gravityscale}), if we impose that
$R_5(t_3)$ in which the gauge fields propagate is the largest extra
dimension
\[ R_{i}<R_5(t_3)\approx \frac{1}{\alpha _{W}M_{f}} \] 
where the approximate equality is due to Eq.(\ref{5dpertbound}), one sees 
that
\[ M_{f}>\bar{M}_{pl}\alpha _{W}^{n/2}\approx 1.5\times
10^{13}\textrm{GeV}.\] 
To restate this, if the gauge fields come from dimensional reduction 
of the {\em largest compact}\/ dimension, perturbativity requires
\[ M_{f}>10^{13}\,{\rm GeV}.\] 
Hence, for this scenario to have \( M_{f}\leq 10^{5}\,\)GeV, there 
must be compact dimensions larger than the dimension in which the gauge 
fields propagate.

Then we take, for simplicity, 5 large extra dimensions of length
\( R' \) in which the gauge fields do not propagate, and one
comparatively smaller extra dimension of length \( R_5(t_3)=
\frac{1}{\alpha _{W}M_{f}} \) in which (electroweak) gauge fields do 
propagate. Since in that case,
\[ M_{f}^{2}=\frac{\alpha _{W}\bar{M}_{pl}^{2}}{(R'M_{f})^{5}}\]
we can choose
\[ R'=\frac{2\times 10^{5}}{M_{f}}\left(\frac{10^{5}\,{\rm GeV}}{M_{f}}\right)^{2/5}=2\,{\rm GeV}^{-1}\left(\frac{10^{5}\,{\rm GeV}}{M_{f}}\right)^{2/5}\]
Furthermore, the experimental limit from neutron-antineutron oscillations 
forces $M_f \geq 10^5\,$GeV \cite{benakli}, 
therefore $R'$ must be smaller than about $2\,$GeV$^{-1}$.  
In the spirit of the original motivation for large extra dimensions, namely
in order to ameliorate the hierarchy problem, we take the smallest
consistent value of $M_f$ and set it to $M_f\approx 10^5$ GeV.

What is important to keep in mind from this section is
\eqr{5dpertbound}. Since $\alpha_W(t_3) \equiv
\frac{g_{4}^{2}(t_{3})}{4\pi } \approx 1/31$, the maximum value that
$V(t_3)M_f$ can take is $31$.  Hence, proton decay constraint requires
$\zeta \geq 3/2$ as claimed below \eqr{eq:defzeta}. Furthermore, to
``unsuppress'' the dimension-6 operator, $L M_f \approx 1$ is required
during baryogenesis. This means that if the distance between the
branes $L$ is equal to the time dependent size of the extra dimension
$R_5$, then $\alpha_W \approx 1$ during baryogenesis.  However, as
shown in Fig.\ \ref{fig:branesetup}, if the distance between the
branes can be adjusted independently of the size of the extra
dimension, this need not be true.  Finally, if we require that no K-K
modes of the $SU(2)_L\times U(1)_Y$ gauge group contribute to the
baryogenesis dynamics, we must have $R_5^{-1} > 100\,$.
This leads to the requirement that $\alpha_W(t) > 10^{-3}$.
We will henceforth take $\alpha_W$ to be a free (time-dependent)
parameter, with a value
\begin{equation}
10^{-3} < \alpha_W(t) <1.
\label{eq:couplingbounds}
\end{equation} 

With a setup in which proton decay and other $B$-violating processes are suppressed by the quark-lepton separation, the next question to ask is which processes dominated at the epoch when $L$ was smaller ($L^{-1}\sim M_f$). Proton decay due to massive fermion exchange, which appears in the effective theory by operators suppressed as $e^{-kLM_f}$ with $k\sim 1$, is the most ``difficult'' to suppress at the present epoch, since one requires $M_fL\geq 50$ at the present time up to factors of order 1, and $R$ is bounded from above by $\alpha_W^{-1}$ if SU$(2)_{\rm W}$ is to remain perturbative in the bulk. Decay mediated by local interactions in the $d=5$ effective field theory, requiring the localized SM fermions to propagate through the bulk with a suppression of $e^{-\mu^2L^2}$, might also be problematic if $\mu^2$ is small relative to $M_f^2$, as it must be for a credible field theory model of localization. Here we would require $\mu^2 L^2\geq 50$ today. Nonperturbative effects involving the extra dimensions should also be considered, for example virtual black holes or wormholes connecting the two branes. However, given very reasonable assumptions for the topology of the corresponding instantons, their action is large enough for the effects to be negligible \cite{ahs}. All other $R$-dependent effects turn out generically to be smaller, including the weak instantons which do not cause proton decay but might be important in baryogenesis (Appendix \ref{app:inst}). Then, assuming that both suppression factors saturate the current bound, as $R$ shrinks from its present value the effective operators suppressed by $e^{-\mu^2L^2}$ grow much more quickly than those varying as $e^{-{\rm const.}M_f R}$. At the time of baryogenesis, for $R$ comparable to $M_f$, the dominant processes (all other things being equal) will be those mediated by the SM fermions propagating through the bulk, for which the suppression is $\sim e^{-\mu^2M_f^2}\sim 1$.

The goal now is to see if without an electroweak phase transition, we can 
create enough baryon asymmetry with the dimension-6 operator which is 
unsuppressed.

\section{Baryogenesis}
The defining characteristic of our class of baryogenesis scenarios is
the leading baryon number-violating dimension-6 operator, one example of 
which is written in the gauge eigenstate basis as
\begin{equation}
\Delta L=\frac{v_s \lambda^{anbr}}{M_{f}^{2}}
(Q_{L}^{a})^{T}_{i}C^{-1}(Q_{L}^{n})_{j}(L_{L}^{b})^{T}_{l}C^{-1}
(Q_{L}^{r})_{m}[c_{1}\epsilon ^{ij}\epsilon ^{lm}+c_2 \epsilon^{i l} 
\epsilon^{j m} + c_3 \epsilon^{i m} \epsilon^{j l}]
\end{equation} 
where we have suppressed color indices, \( \lambda ^{anbr} \) are 
family-dependent coefficients, \( \{i,j,l,m\} \) are \( SU(2)_{L} \) 
indices, summation of indices is implicit, and $c_1$, $c_2$ and $c_3$ are 
constant real coefficients. The wave function overlap integral that gives 
rise to $v_s$ will be taken to be $1$ during baryogenesis. Rotating the 
quarks and leptons into mass eigenstates, we have
\begin{equation}
(Q^{a})_{i}=U_{i}^{a\tilde{a}}(\tilde{Q}^{\tilde{a}})_{i}
\end{equation}
\begin{equation}
(L^{a})_{i}=W_{i}^{a\tilde{a}}(\tilde{L}^{\tilde{a}})_{i}
\end{equation}
which corresponds to the usual flavor mixing matrices 
\( V_{\rm CKM}=U_{u}^{\dagger }U_{d} \)
and \( V_{MNS}=W^{\dagger }_{u}W_{d} \). This then gives
\begin{equation}
\Delta L=\frac{\lambda _{ijml}^{anbr}}{M_{f}^{2}}
(\tilde{Q}_{L}^{a})^{T}_{i}C^{-1}(\tilde{Q}_{L}^{n})_{j}
(\tilde{L}_{L}^{b})_{l}^{T}C^{-1}(\tilde{Q}_{L}^{r})_{m}
[c_{1}\epsilon ^{ij}\epsilon ^{lm}+\ldots ]
\end{equation}
where
\begin{equation}
\lambda _{ijlm}^{anbr}=\lambda ^{a'n'b'r'}U_{i}^{a'a}U_{j}^{n'n}U_{m}^{r'r}W_{l}^{b'b}
\end{equation}
Note that due to the introduction of the dimension-6 operators, the
global symmetry of the theory (in the limit that the Yukawa
couplings vanish) is different from the usual low energy analysis of
the Standard Model. Hence, we have more than one rephasing invariant
CP-violating complex parameter.

\begin{figure}
{\par\centering \resizebox*{.8\columnwidth}{!}{\includegraphics{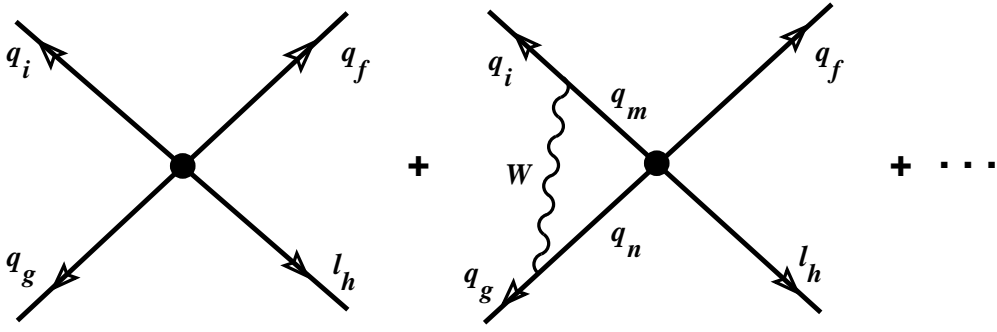}} \par}
\caption{\label{fig:interf} Diagrams interfering for baryogenesis.}
\end{figure}

Now consider for example the interference between the tree level and
the one-loop diagrams shown in Fig.~\ref{fig:interf}, assuming for 
simplicity that $c_3=c_1=0$ and $c_2=1$.  If the tree level 
amplitude is denoted as ${\cal M}_1$ while the one loop amplitude is
labeled as ${\cal M}_2$, we know that ${\cal M}_1$ can be written as
\begin{equation}
{\cal M}_1= R_1 \lambda_{1122}^{atbs}
\end{equation}
where $R_1$ is purely real, while ${\cal M}_2$ will in general be of
the form
\begin{equation}
{\cal M}_2 = (R_2+ i I_2) V_{is}V_{tj}^{*} \lambda_{1122}^{ajbi} 
\end{equation}
where $R_2$ and $I_2$ are real functions.  Note that $I_2$ will in
general arise if we can cut the propagators in the loop (to make two
tree level graphs) and put them on shell simultaneously.  For the $CP$
conjugate reaction ${\cal M}_{\bar{1}}=R_1 \lambda_{1122}^{atbs*}=
{\cal M}_{1}^*$, but 
\begin{equation}
{\cal M}_{\bar{2}}= (R_2+ i I_{\bar{2}}) V_{is}^{*}V_{tj}
\lambda_{1122}^{ajbi*}\neq {\cal M}_2^*
\end{equation}
where in particular, the imaginary part arising from the cut propagators 
is not the complex conjugate of that for the $CP$-conjugate reaction, 
since it derives from kinematics alone.

In the Boltzmann equations for the baryon number density, the probability 
of interest will be
\begin{eqnarray}
D\equiv \langle |{\cal M}_{q l \rightarrow \bar{q} \bar{q}}|^2 \rangle
 - \langle |{\cal M}_{\bar{q} \bar{l} \rightarrow q q}|^2\rangle & = &
2\,{\rm Re}\,[ \langle {\cal M}_1 {\cal M}_2^* - {\cal M}_{\bar{1}} {\cal
M}_{\bar{2}}^* \rangle ] \\ & = & 4 \langle R_1  (I_2
+I_{\bar{2}}) \rangle  \delta  
\end{eqnarray}
where we have defined
\begin{equation}
\delta \equiv  \,{\rm Im}\,[ V_{is}^*V_{tj}
\lambda_{1122}^{ajbi*} \lambda_{1122}^{atbs}]
\end{equation}
and the $\langle \rangle$ represents spin averaging.
Written out explicitly, this interference term is
\begin{equation}
D = \frac{-g^{2}}{4\pi M_{f}^{4}}\sum _{ij}\,
{\rm Im}\,\{\lambda_{1122}^{ajbi*}\lambda^{atbs}_{1122}V_{is}V_{tj}^{*}\}
I_{stij}^{\alpha \beta}\,{\rm Re}\,\{{\rm Tr}\,[\feynslash{p_3}\gamma^\nu 
\gamma_\alpha \feynslash{p_1} \feynslash{p_4} \gamma_\nu \gamma_\beta 
\feynslash{p_2} P_R] \} + \mbox{parity odd} 
\end{equation}
where the external momenta $p_i$ are labeled clockwise and constrained
to satisfy $p_1 + p_2 -p_3 -p_4 =0$, $P_R=(1+\gamma_5)/2$, and
\begin{equation}
I^{\alpha \beta}_{stij} \equiv \,{\rm Re}\,\left\{ 
2 \pi \int \frac{d^4 k}{(2 \pi)^4} 
\left[\frac{(p_3-k)^\alpha}{(p_3-k)^2 -m_{u_{i}}^2}\right] 
\left[\frac{(p_4 +k)^\beta}{ (p_4 +k)^2 -m_{d_{j}}^2 } \right]
\frac{1}{k^2-M_W^2} \right\}
\end{equation}
Note that $I^{\alpha \beta}_{stij}$ vanishes only if the intermediate
quark states cannot go on shell.  If the intermediate quark states
can go on shell, the expression becomes
\begin{equation}
I_{stij}^{\alpha \beta}=-8 \pi^3 \int \frac{d^3 q_1}{(2 \pi)^3 2 E_{q_{1}}}
\frac{d^3 q_2}{(2 \pi)^3 2 E_{q_{2}}} (2 \pi)^4 \delta^{(4)}(q_1 +q_2
-(p_1 +p_2)) \frac{q_1^{\alpha} q_2^{\beta}}{(p_3-q_1)^2 -M_W^2}. 
\end{equation}

\begin{figure}
{\centering \includegraphics{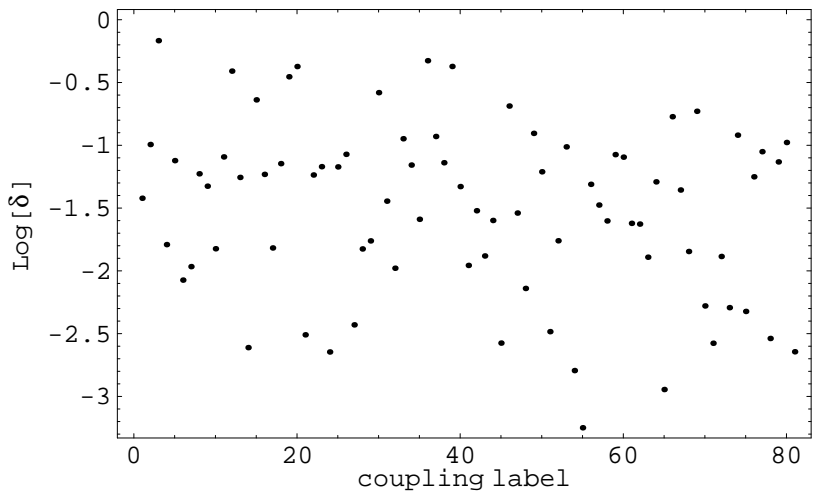} \par}
\caption{A typical distribution of $CP$ violating coefficients from a
random texture.}
\label{fig:imaginarycoeffs}
\end{figure}

Now, let us make some estimates.  Consider the standard
parameterization of \( V_{\rm CKM} \).  Suppose we choose the angles as \(
\theta _{1}=3.366 \), \( \theta _{2}=0.0307 \), \( \theta _{3}=0.0067
\), and a phase \( \delta _{\rm CKM}=1.4633 \) roughly
consistent with experimentally known bounds. We specify in part a Yukawa
texture by choosing the rotation matrix
\[ U_{u}=\exp\left[ i\left(
\begin{array}{ccc} 1 & 0.5 & 0.9\\ 0.5 & 1 & 0.6\\ 0.9 & 0.6 & -2
\end{array}\right)\right]\]
and suppose $U_{i}=W_{i}$.
With a array of random (real) numbers of order 1 for the 
$\lambda ^{anbr}$, we find the effective $CP$ violating imaginary
coefficient to be distributed as shown in Fig.~\ref{fig:imaginarycoeffs}
where we have summed over the indices $i,j$ as if rest of the
transition probability did not depend on $i$ and $j$. As expected,
the quantity $\delta$ (analogous to the Jarlskog invariant in the 
presence of the dimension-6 operators) is around 
$0.1$, much larger than the Jarlskog invariant of $\sim 10^{-5}$. 
We expect similar results if the $\lambda ^{anbr}$ are taken complex
with random phases.

As for the $I^{\alpha \beta}_{stij}$ we have when the Mandelstam
invariant $s\ll M_W^2$ 
\begin{equation}
I^{\alpha \beta}_{stij} \sim 10 g^{\alpha
\beta}\frac{1}{M_W^2} \left(s-(m_{u_{i}}+m_{d_{j}})^2\right)
\end{equation}
while for the top quark, 
\begin{equation}
I_{stij}^{\alpha\beta} \sim 30.
\end{equation}
Note that $M_W$ can be time dependent since 
\begin{equation}
M_W \approx \sqrt{2 \pi \alpha_W} v
\end{equation}
and $\alpha_W$ can be time dependent (anywhere between $0<\alpha_W
\leq 1$). Although $v$ may be time-dependent, we will take it to
be approximately constant \footnote{It is typically somewhere between
$70$ and $180$ GeV for the standard thermal scenario.}.

Now, let us write down the Boltzmann equation for our system of quarks
and leptons.  Denoting the phase space density of particles in an FRW
cosmology with scale factor $a$ as $f_x$, we have
\begin{eqnarray}
\frac{\partial f_x}{\partial t} - \frac{\dot a}{a} \frac{|\vec p|^2}{E_x}
\frac{\partial f_x}{\partial E_x} & = & \sum_{i,j,b,c,d}\int 
\left\{-f_x f_b |{\cal M}_1|^2 - 
2 f_xf_b\,{\rm Re}\,\{{\cal M}_1{\cal M}_2^* \} + 
\frac{f_cf_d}{f_c^{eq} f_d^{eq}} f_x^{eq} f_b^{eq} |{\cal M}_1|^2
\right. \nonumber \\
&  & -2 \frac{f_c f_d}{f_c^{eq} f_d^{eq}} f_x^{eq} f_b^{eq} 
\,{\rm Re}\,\{{\cal M}_1 {\cal M}_2^*\} \nonumber \\
&  & - f_xf_b |{\cal M}_3|^2 - 
2 f_x f_b \,{\rm Re}\,\{{\cal M}_3 {\cal M}_4^* \} + 
\frac{f_i f_j}{f_i^{eq} f_j^{eq}} f_x^{eq} f_b^{eq} |{\cal M}_3|^2
\nonumber \\
&  & \left. -2 \frac{f_i f_j}{f_i^{eq} f_j^{eq}} f_x^{eq} f_b^{eq} 
\,{\rm Re}\,\{ {\cal M}_3 {\cal M}_4^*\} \right\} +\ldots
\end{eqnarray}
where 
\begin{equation}
\int \equiv \frac{1}{2 E_x} \int \prod_{n \in \mbox{\rm reactants}} 
d\,\Pi_n (2 \pi)^4 \delta^{(4)}
\left(p_x+ \sum_{r\in \mbox{\rm reactants}} p_r\right),
\end{equation}
is an operator generating an appropriate momentum integral (with the
measure $d\,\Pi_n= d^3p_n/[2 E_n (2 \pi)^3]$), ${\cal M}_1$ is the
amplitude for the tree level process $x b \rightarrow c d$, and ${\cal
M}_2$ is associated with the one loop correction with intermediate
states $i j$. The amplitude ${\cal M}_3$ is associated with the tree
level amplitude of $x b \rightarrow i j$, where $i j$ are the
intermediate states in the calculation of ${\cal M}_2$, and the one
loop correction to ${\cal M}_3$ involving the intermediate states $c
d$ (appearing in ${\cal M}_1$) is denoted as ${\cal M}_4$.  Here, we
have neglected the Pauli blocking factors for the Boltzmann collision
integral.  Although this is easily generalizable to include the Pauli
blocking factors and although it is not justified when calculating an
accurate number density evolution, the neglect is justified when
calculating an upper bound on the baryon numbers produced.  Hence, for
clarity of physics we shall drop them in this paper.  Because only the
amplitudes with the fermion propagators cut in ${\cal M}_2$ and ${\cal
M}_4$ contribute, one can show that within the collision integral sum
that $ \,{\rm Re}\, \{ {\cal M}_1 {\cal M}_2^*\}= - \,{\rm Re}\,\{
{\cal M}_3 {\cal M}_4^* \} $. Note that the collision integral
vanishes in equilibrium. (A related work on this kind of Boltzmann
equation may be found in Ref.~\cite{dolgov}.)

Thus one arrives at the baryon number density evolution equation
\begin{equation}
\frac{1}{a^3}\frac{d(n_B a^3)}{dt} = 2 \int' \sum_{a,b,i,j,c,d} b_a 
f_a^{eq} f_b^{eq} \,{\rm Re}\, \{ {\cal M}_1 {\cal M}_2^* 
(ab\rightarrow ij \rightarrow cd) \}
\left[ \frac{f_i f_j + f_{\bar i} f_{\bar j}}{f_i^{eq} f_j^{eq}} -
\frac{ f_{\bar{c}} f_{\bar{d}}+ f_c f_d}{f_c^{eq} f_d^{eq}} \right]
\label{eq:genericbaryonevolve}
\end{equation}
where $b_a$ is the baryon number associated with reactant $a$ and the
prime on the integral denotes that all the momenta are to be integrated
over. Note that for each baryon number associated with reactant $a$ to
contribute to the baryon asymmetry, the intermediate and the final
state must be out of equilibrium and not the initial reactant.  This
is expected from unitarity and CPT which implies that
\cite{kolbwolfram}
\begin{equation}
\sum_j | {\cal M}(i\rightarrow j)|^2 = 
\sum_j | {\cal M}(\bar{i} \rightarrow j) |^2
\label{eq:cptunitarity}
\end{equation}
where $j$ sums over all final states: when the probability is summed
over all final states, there cannot be any $CP$-violating enhancement or
subtraction of the probability that survives.  Also, note that
although na\"\i vely one might think that the RHS of 
Eq.~(\ref{eq:genericbaryonevolve}) would vanish, since the quantity inside 
the square bracket is odd under the exchange $i j \leftrightarrow c d$ and 
the summation is over all $i,j,c,d$, there is no cancellation because 
${\rm Re}\, \left\{ {\cal M}_1 {\cal M}_2^* (ab\rightarrow ij \rightarrow cd) 
\right\}$ is also odd under the exchange $i j \leftrightarrow c d$. 
Finally, note that although we have only written down the scattering 
processes explicitly, the decay and inverse decay reactions have an 
analogous form.

Now, instead of accounting for the possible contributions to the
baryon asymmetry from all the reactants together, let us put bounds
on the the maximum contribution that arises from each of the reactants.
As explained in Appendix \ref{appendixboltzmann}, we can derive
from \eqr{eq:genericbaryonevolve}
\begin{equation}
n_{B}\leq \frac{2\eta }{a^{3}}\int dt\,\Gamma _{\rm BV}(t)a^{3}n_{f},
\label{eq:rigoroustext}
\end{equation}
where
\begin{equation}
\eta \equiv \alpha _{W}\sum _{ij} I \,{\rm Im}\,\{\lambda ^{ajbi}\lambda
^{*atbs}V_{is}V^{*}_{tj}\}\sim 0.02 \frac{\bar{T}^2}{ v^2},
\end{equation}
$\bar{T}^2 \approx T_{init}^2 + m_{init}^2 - (m_{ui} + m_{dj})^2$,
$T_{init}$ and $m_{init}$ are temperature and mass scales,
respectively, of the initial state (or final state) reactants,
$m_{ui}$ and $m_{dj}$ are the masses of the particles in the loop, \(
\Gamma _{\rm BV} \) is the tree-level baryon number violating reaction
rate which can be read off from Table \ref{tab:bvrates}. In the Table
the function $W$ defined as
\begin{equation}
W(m_{Q_u^a},T)\equiv  \frac{m_{Q_u^a}^2 T}{2 \pi^2
n_{Q_u^a}^{eq}}K_1(m_{Q_u^a}/T)\approx \frac{1}{\sqrt{1+4 (T/m_{Q_u^a})^2}} ,
\end{equation}
is a time dilation function arising from the thermal averaging and 
$F(m_{Q}, m_{L}, T_Q, T_L)$ is an function of order 1 also arising from 
thermal averaging ($K_1$ is a modified Bessel function of the second
kind).
\begin{table}
{\centering \begin{tabular}{|c|c|c|c|}
\hline 
\( \Gamma _{\rm BV} \) & reaction & magnitude & \( \Delta (B+L) \) \\
\hline 
\( \Gamma _{\rm BV\ dec}(1+\epsilon ) \)&
\( Q_{u}^{a}\rightarrow \bar{L}_{d}^{b}\bar{Q}_{d}^{s}\bar{Q}_{u}^{t} \)&
\( \lambda^2\left( \frac{m_{Q_{u}^{a}}}{M_{f}}\right) ^{4}\frac{m_{Q_{u}^{a}}}{(2\pi )^{3}32}(1+\epsilon ) W(m_{Q_u^a},T_Q)\)&
\( -2 \)\\[1.5mm]
\hline 
\( \Gamma _{\rm BV\ dec}(1-\epsilon ) \)&
\( \bar{Q}_{u}^{a}\rightarrow L_{d}^{b}Q_{d}^{s}Q_{u}^{t} \)&
\( \lambda^2\left( \frac{m_{Q_{u}^{a}}}{M_{f}}\right) ^{4}\frac{m_{Q_{u}^{a}}}{(2\pi )^{3}32}(1-\epsilon ) W(m_{Q_u^a},T_Q)\)&
\( 2 \)\\[1.5mm]
\hline 
\( \Gamma _{\rm BV\ scatt}(1+\epsilon ) \) & \rule{0cm}{7.5mm}
\( L_{d}^{b}Q_{u}^{a}\rightarrow \bar{Q}_{d}^{s}\bar{Q}_{u}^{t} \) &
\( \frac{n_{L,Q}\lambda^2(m_{L^{b}_{d}}^{2}+m_{Q_{u}^{a}}^{2} + T_Q^2+T_L^2)}{16\pi M_{f}^{4}}(1+\epsilon ) F(m_{Q_u^a}, m_{L^{b}_{d}}, T_Q, T_L) \)&
\( -2 \)\\[1.5mm]
\hline 
\( \Gamma _{\rm BV\ scatt}(1-\epsilon ) \) & \rule{0cm}{7.5mm}
\( \bar{L}_{d}^{b}\bar{Q}_{u}^{a}\rightarrow Q_{d}^{s}Q_{u}^{t} \)&
\( \frac{n_{\bar{L},\bar{Q}}\lambda^2(m_{L_{d}^{b}}^{2}+m_{Q_{u}^{a}}^{2}+T_Q^2+T_L^2)}{16\pi M_{f}^{4}}(1-\epsilon )F(m_{Q_u^a}, m_{L^{b}_{d}}, T_Q, T_L) \)&
\( 2 \)\\[1.5mm]
\hline 
\end{tabular}\par}
\caption{\label{tab:bvrates}
The dominant reaction rates. The upper indices on
leptons \protect\( L\protect \) and quarks \protect\( Q\protect \) label family
number while the subscript indices label denote up-type or down-type
({\em e.g.}\/\ \protect\( L_{u}\protect \) is a neutrino). 
All other quantum numbers
are implicit. The small number \protect\( \epsilon \protect \) indicates the
splitting due to the $CP$-violating phases. Only the quark decay rates are 
displayed, as they will be the most relevant quantities. The coupling 
\protect\( \lambda \protect \) of the dimension-6 operator is assumed to be of 
order 1.}
\end{table}
Using these and other results presented in Appendix \ref{appendixboltzmann}, 
we can place an upper bound on the baryon asymmetry generated by any
single species out of equilibrium as
\begin{equation}
\label{eq:generalbound}
\frac{n_{B}}{s} < 2\eta \left[\frac{\Gamma _{\rm BV}(t)}{\Gamma _{\rm tot}(t)}
\right] _{\rm max}
\end{equation}
where the maximum is taken at any time during baryogenesis. Note that this 
bound is independent of cosmology (without specifying the time dependence of 
the scale factor \( a \)) as long as the usual Boltzmann equations are valid. 

Let us first consider the quarks. Instead of writing 
\( \Gamma _{\rm tot} \) as the sum of rates for all the channels, it is 
more illuminating to consider separately the cases when \( \Gamma _{\rm tot}\) 
is dominated by either the baryon number-conserving annihilation rate or the 
decay rate. The $B$-conserving annihilation rate is dominated by a t-channel 
quark annihilation into gluons, with the possible exception of the top quark 
for which the rate can be dominated by a weak scattering. Explicitly, we use 
\begin{equation}
\Gamma _{\rm ann}\approx n_{Q}\frac{\alpha _{S}^{2}}{m_{Q}^{2}+T_Q^2}
F_2(m_Q, T_Q) \end{equation} 
\begin{equation}
\Gamma _{t-{\rm ann}}\approx \Gamma _{\rm ann}
\left(\frac{m_{t}}{M_{W}}\right)^{4}
\left(\frac{\alpha _{W}}{\alpha _{S}}\right)^{2}
|V_{\rm CKM}|^{2}\equiv \Gamma _{\rm ann}/E_{t}
\end{equation}
where \( |V_{\rm CKM}|^{2} \) is the appropriate quark mixing matrix
element squared, and $F_2$ is a function encoding the effect of thermal 
averaging, similar to $F$. The $B$-conserving decay rate is dominated 
by a weak decay rate, which is enhanced for the top quark due to the 
longitudinal \( W \) component going on shell. We use
\begin{equation}
\Gamma _{\rm weak\ dec}\approx \frac{\alpha _{W}^{2}}{16\pi }
\left(\frac{m_{Q}}{M_{W}}\right)^{4}m_{Q}|V_{\rm CKM}|^{2} W(m_{Q_u^a},T_Q)
\end{equation}
\begin{equation}
\Gamma _{t-{\rm dec}}\approx \Gamma _{\rm weak\ dec}
\frac{\pi }{\alpha _{W}}\left(\frac{M_{W}}{m_{t}}\right)^{2}\equiv 
\Gamma _{\rm weak\ dec}/F_{t}
\end{equation}
Explicitly, \( E_{t}\approx \frac{4}{|V_{\rm CKM}|^{2}}\vscale^4 \) while
\( F_{t}\approx 0.05 \vscale^{-2} \), which implies that the weak decay rate is
not much enhanced for the top quark through the Goldstone emission. Given these
rates, we list the leading bound \( n_{B}/s \) ({\em i.e.}\/\ the upper bound 
on the baryon fraction relevant to the case that the given reactions are 
the dominant contributions) in Table \ref{tab:quarkrates}. 
The lepton sector is similar except with $\alpha_s 
\rightarrow \alpha_{em}$ and no top quark enhancements. In addition, the 
leading number density changing channel for the neutrinos may be 
coannihilations instead of self-annihilations.
\begin{table}
{\centering \begin{tabular}{|c|c|c|}
\hline 
ratio \#&
\( \Gamma _{\rm BV}/\Gamma _{\rm tot} \)&
\( (n_{B}/s)_{\rm max} \)\\
\hline 
1&
\( \Gamma _{\rm BV\ dec}/\Gamma _{\rm ann} \)& \rule{0cm}{7mm}
\( \frac{2 \eta}{\alpha
_{S}^{2}}\frac{m_{Q}^4}{M_{f}^4} \left[ \frac{m_{Q} (m_Q^2+T_Q^2)W(m_Q, T_Q)}{{n_{Q}} F_2(m_Q, T_Q)} \right]_{\rm max}(\times E_{t})\)\\[2mm]
\hline 
2&
\( \Gamma _{\rm BV\ scatt}/\Gamma _{\rm ann} \)& \rule{0cm}{7mm}
\( \frac{\eta}{8\pi \alpha _{S}^{2}}\frac{(m_{Q}^{2}+T_Q^2)(m_{f_{2}}^{2}+m_{Q}^{2}+T_Q^2)}{M_{f}^{4}}\left[ \frac{n_{f_{2}}(t)F(m_Q, m_{f_2}, T_Q)}{n_{Q}(t)F_2(m_Q, T_Q)}\right] _{\rm max}(\times E_{t}) \)\\[1.5mm]
\hline 
3&
\( \Gamma _{\rm BV\ dec}/\Gamma _{\rm weak\ dec} \)& \rule{0cm}{6.5mm}
\( \frac{2 \eta}{g^{4}}(\frac{M_{W}}{M_{f}})^{4}\frac{1}{|V_{\rm CKM}|^{2}}(\times F_{t}) \)\\[1mm]
\hline 
4&
\( \Gamma _{\rm BV\ scatt}/\Gamma _{\rm weak\ dec} \)& \rule{0cm}{7mm}
\( \frac{32\pi \eta}{g^{4}}(\frac{M_{W}}{M_{f}})^{4}\frac{(m_{f_{2}}^{2}+m_{Q}^{2}+T_Q^2+T_{f_2}^2)}{m_{Q}^{2}
|V_{\rm CKM}|^{2}}\left[\frac{n_{f_2}F(m_Q, 
m_{f_2}, T_Q, T_{f_2})}{m_Q^3W(m_Q, T_Q)}\right]_{\rm max}(\times F_{t}) \)
\\[1.5mm]
\hline 
\end{tabular}\par}
\caption{\label{tab:quarkrates}\protect\( n_{B}/s\protect \) for quarks. 
For the top quark, each of these ratios must be multiplied by 
\protect\( E_{t}\protect \) or \protect\( F_{t}\protect \). 
However, these extra factors do not change our conclusions.}
\end{table}
From the last section, we have \( M_{f}>100 \) TeV which implies
\begin{equation}
\eta (M_{W}/M_{f})^{4} < 3\times 10^{-13} \vscale^2.
\end{equation}
Finally, we need to constrain the number density of the particles. Since the
elastic scattering rates are at least as large as the total inelastic 
scattering rates, to absolutely forbid any phase transitions, we restrict 
the number density of particles to be smaller than the thermal density at 
\( T_{*}=30 \) GeV, 
{\em i.e.}
\begin{equation}
\label{eq:densitylimit}
n^{\rm max}_{Q,L}<0.2T_{*}^{3}
\end{equation}
if relativistic or
\begin{equation}
n^{\rm max}_{Q,L}<0.13(m_{Q,L}T_{*})^{3/2}e^{-m_{Q,L}/T_{*}}
\end{equation}
if nonrelativistic \footnote{Note that no color factors were included.}.

Now, consider the case when the branching ratio is approximated by ratio 3 in
Table \ref{tab:quarkrates}. For the top quarks out of equilibrium, we have 
\begin{equation}
\mbox{ratio 3}=5 \times 10^{-15}\vscale^2
\end{equation}
while for all other reactants, we have
\begin{equation}
\mbox{ratio 3} <6 \times 10^{-17} \left( \frac{1}{ |V_{\rm
CKM}|^2} \right)\vscale^2  
\label{eq:decotherquarks}
\end{equation}
where we have used the bottom quark mass of about 4.3 GeV.  For the
top quark, because the upper bound here is smaller than what is needed
for sufficient baryogenesis and since the same numerator appears in
ratio 1, the $B$-violating decay channel is ruled out for the top
quark. For the bottom quark, since the leading decay channel has \(
|V_{cb}|^{2}\approx 1.6\times 10^{-3} \), \eqr{eq:decotherquarks}
also rules out sufficient baryogenesis.  All other
channels also fail the tests under ratios 1 and 3 which means that the
$B$-violating decay channel cannot help in generating sufficient
baryon asymmetry.

Consider now ratios 2 and 4. Ratio 4 requires for the top quark that
\begin{equation}
\frac{n_B}{s} < 4\times 10^{-15}
\end{equation}
if we assume that $F\sim W \sim 1$.  Hence, ratios 3 and 4 completely
rule out the top quark as a significant contributor to the baryon
asymmetry. 

Let us now consider the scattering of the out-of-equilibrium particle $f$ 
with every other particle labeled as $f_2$ in view of ratio 2. It can easily 
be seen that ratio 2 is not very constraining, as it can in principle be
made arbitrarily large. Hence, we can consider a more stringent version of
the entropy bound corresponding to ratio 2, \eqr{eq:rigoroustext} which we 
rewrite as
\begin{equation}
\label{eq:bbound}
\frac{n_B}{s} \leq \frac{ 2 \eta (m_f^2 + m_{f_2}^2 + \tbarsq)}{M_f^4 s}
\frac{1}{a^3} \int_{t_B}^{t_f} dt\,a^3 g_1(m_f, m_{f_2}, T_f,T_{f_2}) 
n_f n_{f_2}
\end{equation}
where $s$ is the entropy density, $T_i$ is the temperature of particle
species $i$, $\bar{T}_i$ is the characteristic temperature of species
$i$ inside the integral, $g_1$ is a kinematic function of order 1,
$t_B$ marks the beginning time of baryogenesis, and $t_f$ marks the
end time of baryogenesis. Instead of using the maximum contribution
to the entropy, it is sufficient to consider any one contribution
since we are placing an upper bound. Since we are guaranteed that the
reaction changing the number density of $f$ is out of equilibrium, and
there exists a contribution to the entropy from the weak interaction 
($f f_2 \rightarrow \mbox{final products}$) with the matching integral of
the form $\int_{t_B}^{t_f} dt\, a^3 n_f n_{f_2}$ (since $f$ is necessarily 
out of equilibrium), we can write using Eq.~(\ref{eq:entropyboltzmann})
\begin{equation}
\label{eq:weakentropy}
s \geq \frac{\alpha_W^2 |V_{\rm CKM}|^2 (m_f^2+m_{f_2}^2+\tbarsq)}{M_W^4} \frac{1}{a^3} 
\int_{t_B}^{t_f} dt\, a^3 g_2(m_f, m_{f_2}, T_f,T_{f_2}) n_f n_{f_2}
\end{equation}
where $g_2$ is another order 1 dimensionless function. It is
important to note that we have assumed the kinematic viability of the
weak interaction in Eq.~(\ref{eq:weakentropy}). Combining
\eqr{eq:weakentropy} and \eqr{eq:bbound}, we have
\begin{equation}
\label{eq:nogo1}
\frac{n_B}{s} \leq \frac{2 \eta}{\alpha_W^2 |V_{\rm CKM}|^2}
\left(\frac{M_W}{M_f}\right)^4 \approx 5 \times 10^{-13}\frac{1}
{|V_{\rm CKM}|^2} \vscale^2
\end{equation}
independently of $\alpha_W$.  Hence, with $|V_{\rm CKM}|^2 \leq
10^{-3}$, sufficient baryon asymmetry generation seems feasible.  A
case in which \eqr{eq:nogo1} applies is when out of equilibrium $b$
quarks are interacting with the $c$ quarks at an effective temperature
of 30 GeV.  In particular, if the baryon number conserving weak
interaction channel is dominated by $b c \rightarrow c s$ which gives
the $|V_{cb}|^2\approx 1.6\times 10^{-3}$ suppression, sufficient
baryogenesis in this scenario seems possible.  We will return to this
possibility later and constrain it further.

Another possibility to evade \eqr{eq:nogo1} is to have the weak
interaction be kinematically suppressed.  In the case of out of
equilibrium $c$ quark, the reaction $ c \nu_{\tau} \rightarrow s \tau$
has the kinematic suppression necessary to evade \eqr{eq:nogo1} in the
limit of zero temperature since $m_\tau \approx 1.7\,$GeV while
$m_c\approx 1.5\,$GeV.  In Table \ref{tab:possible}, we list all the
weak interactions that are kinematically suppressed such that
sufficient baryogenesis is possible with respect to the constraints
thus far.

\begin{table}
\begin{tabular}{|c|c|c|c|c|}
\hline 
out of equilibrium particle&
weak interactions&
Threshold Temp.\ (GeV)&
Relevant&
e.g.\ BV\\
\hline
\hline 
\( c \)&
\( c\bar{\nu }_{\tau }\rightarrow s\bar{\tau } \)&
\( 0.4 \)&
Y&
\( c\nu _{\tau }\rightarrow \bar{d}\bar{d} \)\\
\hline 
\( s \)&
\( s\nu _{\tau }\rightarrow u\tau  \)&
\( 1.5 \)&
Y&
\( s\nu _{\tau }\rightarrow \bar{d}\bar{u} \)\\
\hline 
\( d \)&
\( d\nu _{\tau }\rightarrow u\tau  \)&
\( 1.7 \)&
Y&
\( d\nu _{\tau }\rightarrow \bar{d}\bar{u} \)\\
\hline 
{}''&
\( d\nu _{\mu }\rightarrow u\mu  \)&
\( 0.1 \)&
N&
\\
\hline 
\( u \)&
\( u\bar{\nu }_{\tau }\rightarrow d\bar{\tau } \)&
\( 1.7 \)&
Y&
\( u\nu _{\tau }\rightarrow \bar{d}\bar{d} \)\\
\hline 
{}''&
\( u\bar{\nu }_{\mu }\rightarrow d\bar{\mu } \)&
\( 0.1 \)&
N&
\\
\hline 
{}''&
\( u\bar{\nu }_{e}\rightarrow d\bar{e} \)&
\( 0.006 \)&
N&
\\
\hline 
\( e \)&
\( ue\rightarrow d\nu _{e} \)&
\( 0.005 \)&
N&
\\
\hline 
\( \nu _{\tau } \)&
\( \nu _{\tau }\bar{c}\rightarrow \bar{d}\tau  \)&
\( 0.21 \)&
N&
\\
\hline 
{}''&
\( \nu _{\tau }s\rightarrow u\tau  \)&
\( 1.505 \)&
Y&
\( \nu _{\tau }s\rightarrow \bar{d}\bar{u} \)\\
\hline 
{}''&
\( \nu _{\tau }d\rightarrow u\tau  \)&
\( 1.7 \)&
Y&
\( \nu _{\tau }d\rightarrow \bar{d}\bar{u} \)\\
\hline 
{}''&
\( \nu _{\tau }\bar{u}\rightarrow \bar{d}\tau  \)&
\( 1.7 \)&
Y&
\( \nu _{\tau }u\rightarrow \bar{d}\bar{d} \)\\
\hline 
\( \nu _{\mu } \)&
\( \nu _{\mu }d\rightarrow u\mu  \)&
\( 0.1 \)&
N&
\\
\hline 
{}''&
\( \nu _{\mu }\bar{u}\rightarrow \bar{d}\mu  \)&
\( 0.1 \)&
N&
\\
\hline 
\( \nu _{e} \)&
\( \nu _{e}\bar{u}\rightarrow \bar{d}e \)&
\( 0.005 \)&
N&
\\
\hline
\end{tabular}
\caption{\label{tab:possible} The set of reactions evading the
simplest bound. These reactions are suppressed when the temperature 
falls below the threshold temperature.}
\end{table}

According to Table \ref{tab:possible}, some weak interactions are
kinematically suppressed only below the temperature of the QCD phase
transition, which is assumed to occur at around $T\approx 0.2\,$
GeV \footnote{Again, the reader is reminded that by temperature, here
we mean a scale for particle density rather than implying that the
particles are truly in equilibrium.}. These interactions hence really
are not suppressed in the temperature range of interest and the
baryogenesis bound is not evaded for these reactants.  Hence, these
are labeled irrelevant in Table \ref{tab:possible}. Only interactions
involving $\nu_\tau$ then remain as viable candidates for the
kinematically suppressed baryon number conserving channel, if we
assume that its mixing with $\nu_\mu$ or $\nu_e$ is less than
$10^{-4}$.  If the neutrino sector is mixed, all the bounds return and
there are no candidates that can generate sufficient baryon
asymmetry \footnote{By considering the mixing of $\nu_\tau$, we are
assuming that the neutrinos have mass.  In that case, there may
generically be scalars that can produce dimension-5 operators.
However, if these scalars are sufficiently heavy, then their
contribution will be negligible.  We do not analyze such cases, for
which beyond the Standard Model particles actively participate in the
short distance physics.}.

In both cases where \eqr{eq:nogo1} allows sufficient baryogenesis, the
baryon number-conserving weak interactions between the reactants
contribute too little to the entropy to rule out baryogenesis.
In such situations, let us consider whether the self-annihilation
interaction contribution to the entropy is enough to dilute away the
baryon asymmetry contribution.  Start with \eqr{eq:bbound} and again
approximate \( g_{1}\approx 1 \). However, instead of
\eqr{eq:weakentropy}, consider\begin{equation}
\label{eq:selfann}
s\geq \frac{\langle \sigma _{Af_{2}}v\rangle }{a^{3}}\int _{t_{c}}^{t_{f}}dt\,a^{3}n_{f_{2}}^{2}+\frac{\langle \sigma _{Af}v\rangle }{a^{3}}\int _{t_{B}}^{t_{f}}dt\,a^{3}n_{f}^{2}
\end{equation}
which corresponds to the entropy bound arising from self-annihilation
reaction contributions, where the contribution to the entropy from the 
species \( f_{2} \) only begins once it goes out of equilibrium, at time 
\( t_{c} \), while the $f$ self-annihilation contribution persists throughout
the entire duration $[t_B,t_f]$ of baryogenesis.  Here, the thermal
averaged quantities (such as \( \langle \sigma _{Af}v\rangle \)) have
their temperature fixed at some value, characteristic of the
integration interval. The advantage of considering the bound arising 
from self-annihilation reactions is that they change the number density 
and are always possible at tree level if there is a neutrino lighter than 
(or degenerate with) \( \nu _{\tau } \).  We shall assume that the
self-annihilation channel is open \footnote{At the one loop level,
there is a nonlocal contribution to $\nu \bar{\nu}\rightarrow \gamma
\gamma$ that is nonzero even in the zero neutrino mass
limit \cite{crewther}.}.  Suppose first that the \( n_{f_{2}}^{2} \) term
dominates over the second term. Then, using the elementary inequality
\begin{equation} \label{trivineq}
n_{f_{2}}^{2}+n_{f}^{2}\geq 2n_{f}n_{f_{2}}
\end{equation}
we can write
\begin{equation}
\frac{n_{B}}{s}\leq \frac{R}{2M^{2}_{f}\langle \sigma _{Af_{2}}v\rangle }\left[ 1+\frac{\langle \sigma _{A_{f_{2}}}v\rangle }{\langle \sigma _{Af}v\rangle }+\xi \right] 
\end{equation}
where we have defined
\begin{equation}
R\equiv \frac{2 \eta (m_{f}^{2}+m_{f_{2}}^{2}+\bar{T}_f^{2}+\bar{T}_{f_2}^2)}{M_{f}^{2}}
\end{equation}
and
\begin{equation}
\xi \equiv \frac{\int _{t_{B}}^{t_{c}}dt\,a^{3}n_{f_{2}}^{2}}{\int _{t_{c}}^{t_{f}}dt\,a^{3}n_{f_{2}}^{2}}
\end{equation}
which measures how much the $f_2$ particles can contribute to
baryogenesis through interaction with out of equilibrium particle $f$
without contributing to the entropy.  This can be seen from noting
that \( n_{f_{2}} \) is the equilibrium distribution during \( t\in
[t_{B},t_{c}] \) by definition.  Producing baryon asymmetry for a long
time without producing any entropy (i.e. letting $t_c$ approach $t_f$)
is the best that can be done to circumvent the suppression of the
dimension 6 operator.  For sufficient baryogenesis (\( n_{B}/s\simeq 10^{-10} \)), we thus require
\begin{equation}
\label{eq:lemma1}
\xi \geq \frac{\langle \sigma _{A_{f_{2}}}v\rangle }{\langle \sigma _{Af}v\rangle }\left[ \frac{2}{\delta }-1\right] -1
\end{equation}
 where\begin{equation}
\delta \equiv \frac{R}{M_{f}^{2}\langle \sigma _{Af}v\rangle }\times 10^{10}
\end{equation}
If the \( n_{f}^{2} \) dominates 
in Eq.~(\ref{eq:selfann}), one can similarly derive Eq.~(\ref{eq:lemma1}),
thereby showing that it is true in general. Now, noting that one can
write 
\begin{equation}
\int _{t_{B}}^{t_{f}}dt\,a^{3}n_{f}^{2}\leq \delta \int _{t_{B}}^{t_{f}}dt\,a^{3}n_{f}n_{f_{2}}
\end{equation}
and using Eq.~(\ref{trivineq}), one can write
\begin{equation}
(2-\delta )\int _{t_{B}}^{t_{f}}dt\,a^{3}n_{f}n_{f_{2}}\leq \int _{t_{B}}^{t_{f}}dt\,a^{3}n^{2}_{f_{2}}.
\end{equation}
It follows that
\begin{equation}
\label{eq:maineq}
\int _{t_{B}}^{t_{f}}dt\,a^{3}n_{f}n_{f_{2}}\leq \frac{(1+\frac{1}{\xi })}{(2-\delta )}\int _{t_{B}}^{t_{c}}dt\,a^{3}n_{f_{2}}^{2}
\end{equation}
which is useful if \( \xi >1 \) and \( \delta <1 \), since as we
mentioned before, \( n_{f_{2}} \) is the equilibrium distribution
during the time interval \( [t_{B},t_{c}] \). Hence, we shall use
Eq.~(\ref{eq:maineq}) with parameterized cosmology to put a bound
on cosmology from the condition of sufficient baryogenesis.

We parameterize the unknown cosmology as
\begin{equation}
a\propto t^{n},
\end{equation}
\begin{equation}
T\propto a^{-\nu }
\end{equation}
and determine  \( t_{c} \) by the out of equilibrium condition
\begin{equation}
\frac{\dot{a}(t_{c})}{a(t_{c})}= \max[j\langle \sigma
_{Af_{2}}v\rangle n_{f_{2}}(t_{c}), \Gamma_{f_2}]
\label{eq:outofeq}
\end{equation}
where \( j \) is an \( {\mathcal O}(1) \) constant and $\Gamma_{f_2}$
is the decay rate of $f_2$.  Because the ratio of the decay rate
$\Gamma_{f_2}$ to the expansion rate increases as a function of time,
when the decay becomes the dominant chemical potential changing
reaction, it can lead to an out-of-equilibrium number density of the $f_2$ 
{\em only if}\/
the final states of the decay are not in equilibrium.  Hence, the
maximum above applies only if the decay products are out of
equilibrium, and otherwise, only the self-annihilation should be taken
into account.  Furthermore, note that any choice of $t_c$ that we make
in \eqr{eq:outofeq} corresponds to a different choice of cosmology
because $t_c$ sets the scale of expansion rate of the universe.
The larger $t_c$ is, the slower is the expansion rate, and hence the slower 
the dilution rate of any accumulation of baryon asymmetry.  As far as
interpreting the parameter $\nu$ is concerned, it is shown in Appendix
\ref{appendixscaling} that $\nu \leq 1 $ for most physical systems
(even with external sources such as the inflaton) confined to 3
spatial dimensions, unless there exists a cooling mechanism different
from the expansion of the 3 spatial dimensions. However, with the
presence of extra dimensions, the energy exchange with the bulk
exotics may induce a more rapid cooling allowing for $ \nu >1 $.  If
$\langle \sigma _{Af_{2}}v\rangle n_{f_{2}}(t_{c})$ determines the out
of equilibrium time of $f_2$, we find
\begin{equation}
\label{eq:mainnumeq}
\frac{n_{B}}{s} < \frac{\tilde{c}_2 R}{j\tilde{c}M_{f}^{2}\langle
\sigma _{Af_{2}}v\rangle
}\frac{m_{f_{2}}^{2}}{T_{c}^{2}}\frac{(1+\frac{1}{\xi })}{(2-\delta
)}\frac{1}{\nu \pi ^{2}}\int _{T_{c}/T_{B}}^{1}dx\,x^{\frac{1}{\nu
}(\frac{1}{n}+3)-3}\frac{K_{2}^{2}(\frac{m_{f_{2}}}{T_{c}}x)}
{K_{2}(\frac{m_{f_{2}}}{T_{c}})}
\end{equation}
where $\tilde{c}\equiv  2\pi^2 g_{*S}/45$, $g_{*S}$ is the number of
degrees of freedom contributing to the entropy at time $t_c$, and the 
factor of $\tilde{c}_2$ accounts for (the square of) the degrees of 
freedom in the $f_2$ number density ($K_2$ is a modified Bessel 
function of the second kind). 
We will assume that $g_{*S}$ accounts for at least the 16
degrees of freedom of the gluons in the thermal bath.  As for
$\tilde{c}_2$, if $f_2$ for example is a quark, then it will be $9$
for the 3 color degrees of freedom (the 2 helicity states were already
accounted for, which means that if $f_2$ were Weyl neutrinos, this should
be divided by 4). If $\Gamma_{f_2}$ determines the out of equilibrium
time of $f_2$, then this quantity should be multiplied by
\begin{equation}
\frac{\langle \sigma_{Af_{2}} v \rangle m_{f_2}^2 T
K_2(\frac{m_{f_2}}{T})}{\pi^2 \Gamma_{f_2}}.
\label{eq:correction}
\end{equation} 

For the relevant reactions in Table~\ref{tab:possible}, a 
\( \nu _{\tau } \) and a quark are always paired. For the quarks out 
of equilibrium, \( m_{f_{2}}=m_{\nu _{\tau }}\approx 0 \) and \(
m_{f}=m_{q} \): 
\begin{equation} 
\xi \geq \frac{\alpha
_{W}^{2}(\bar{T}_{\nu _{\tau }}^{2}+m_{\nu _{\tau
}}^{2})(m_{q}^{2}+\bar{T}_{q}^{2})}{\alpha _{s}^{2}M_{Z}^{4}}\left[
\frac{2}{\delta }-1\right] -1,
\end{equation}
\begin{equation}
\delta =\frac{2 \eta (m_{q}^{2}+m_{\nu _{\tau }}^{2}+\bar{T}_{q}^2 
+\bar{T}_{\nu _{\tau }}^{2})}{M_{f}^{2}}
\frac{(m_{q}^{2}+\bar{T}_{q}^{2})}{\alpha _{s}^{2}M_{f}^{2}}\times 10^{10}.
\end{equation}
These equations tell us that for all effective temperatures (between
$1.7$ and $0.2\,$GeV) and quark masses (less than $2\,$GeV) relevant
to our scenario, we have the conditions $\delta \ll 1$ and $\xi \gg
1$, which allow nontrivial bounds to come from \eqr{eq:mainnumeq}.
From \eqr{eq:mainnumeq}, it is easy to show that there are no value of
$\nu>0$ and $n>0$ such that sufficient baryon asymmetry is generated.
Hence, there are no scenarios in which there is sufficient
baryogenesis through out of equilibrium quarks interacting with
$\nu_\tau$.

For the case of \( \nu _{\tau } \) out of equilibrium and \( q \) in
equilibrium, \( m_{f_{2}}=m_{q} \) and \( m_{f}=m_{\nu }\approx 0 \)
and we have a similar result:
\begin{equation} 
\xi \geq \frac{\alpha
_{s}^{2}M_{z}^{4}}{\alpha _{W}^{2}(\bar{T}_{q}^{2}+m_{q}^{2})(m_{\nu
_{\tau }}^{2}+\bar{T}_{\nu _{\tau }}^{2})}\left[ \frac{2}{\delta
}-1\right] -1,
\end{equation}
\begin{equation}
\delta =\frac{2 \eta (m_{q}^{2}+m_{\nu _{\tau }}^{2}+\bar{T}_{q}^2+\bar{T}_{\nu _{\tau }}^{2})}{M_{f}^{2}}\frac{M_{z}^{4}}{\alpha _{W}^{2}(m_{\nu _{\tau }}^{2}+\bar{T}_{\nu _{\tau }}^{2})M_{f}^{2}}\times 10^{10}.
\end{equation}
Again, it is easy to check that for all relevant effective
temperatures (between $1.7\,$ and $0.2\,$GeV) and quark masses (less
than $2$ GeV), we have the conditions $\delta \ll 1$ and $\xi \gg 1$
which allow nontrivial bounds to come from \eqr{eq:mainnumeq}.  As
before, \eqr{eq:mainnumeq} rules out the possibility of sufficient
baryogenesis for $n>0$ and $\nu>0$.  Hence, even neglecting the
mixings, it seems nearly impossible to have sufficient baryogenesis
with $\nu_\tau$ through the dimension-6 operator.

\begin{figure}
{\centering \includegraphics{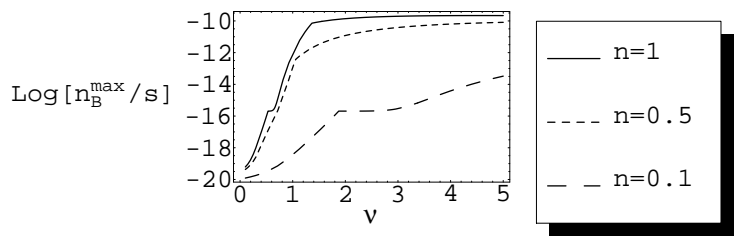} \par}
\caption{\label{fig:almostenough} 
A conservative upper bound for the baryon asymmetry generated by 
out of equilibrium $b$ quarks interacting with $c$ quarks. 
The parameters $n$ and $\nu$ parameterize various possible nonstandard 
cosmologies. A value $n>1$ corresponds to inflating cosmologies and 
in the absence of abnormal cooling $\nu \leq 1$. This scenario does 
not generate enough baryon asymmetry.}
\end{figure}

Let us now return to the case where the baryon number conserving weak
interaction channel is dominated by $b c \rightarrow c s$ which was
seen to have sufficient suppression for us to further consider the
possibility of successful baryogenesis. In applying the reasoning that 
led to \eqr{eq:mainnumeq}, we can improve the bounds compared to just
\eqr{eq:mainnumeq} by considering that the entropy also has a lower
bound at temperature $T_c$ of
\begin{equation}
s(T_c) a^3(T_c) > n_g(t_B) a^3(t_B) + n_{f2}(t_B) a^3(t_B)
\end{equation}
where $n_g$ corresponds to the density of gluons and by putting $R$
inside the time integral and tying all the temperature scales
together. The bound then becomes (similarly to before) the smaller of
either
\begin{equation}
\frac{n_B}{s} <  \frac{\tilde{c}_2}{\nu M_f^2 \langle \sigma_{Af_{2}}v \rangle}
\frac{m_{f_{2}}^2}{M_f^2} 
\left( \frac{T_c}{T_B} \right)^{-3/ \nu} \!\!
\int_{\frac{T_{c}}{T_{B}}}^{1} dx \, 
x^{\frac{1}{\nu}(\frac{1}{n}+3)-3}
\frac{(m_f^2 + m_{f_{2}}^2 + 2 T_c^2/x^2)T_c \eta(x)
K_{2}^{2}(\frac{m_{f_{2}}}{T_{c}} x)}{K_2(\frac{m_{f_{2}}}{T_c})(
\sqrt{\tilde{c}} 
m_{f_{2}}^2 T_B K_2(\frac{m_{f_{2}}}{T_B}) + \pi^2 n_g[T_B])}
\label{eq:alternate}
\end{equation}
or \eqr{eq:mainnumeq}.  The large enhancement for the $b$-$c$
interaction scenario in both \eqr{eq:mainnumeq} and \eqr{eq:alternate}
comes from the long time $t_c$ during which baryogenesis occurs while
$f_2$ is in equilibrium.  This leads to an accumulation of any net gain 
in baryon asymmetry over the entropy.  More explicitly, $t_c$ varies as
$1/K_2(m_{f_{2}}/T_c)$ (or the equivalent factor from 
\eqr{eq:correction} when appropriate) which comes from the
exponentially large ratio of the short distance physics scale to the
expansion rate scale when $f_2$ goes out of equilibrium.  The
expansion rate must be exponentially small when $f_2$ goes out of
equilibrium (if at all) because the self-annihilation reaction must
maintain equilibrium even in the nonrelativistic regime in which the
number densities are exponentially suppressed.  The resulting bound
with $m_f= m_b= 4.3\,$GeV and $m_{f_2}=m_c=1.5\,$GeV is shown in
Fig.~\ref{fig:almostenough}.  As one can see, even with such an
enhancement, there is insufficient baryogenesis (or marginally enough)
in most cases. Even for the marginal cases, we do not expect the
scenario to be realizable given the conservative nature of the bound.
Other scenarios with small $|V_{CKM}|^2$ weak interactions while the
heavier particle is out of equilibrium ({\em i.e.}\/\ out of equilibrium 
$b$ interacting with $u$, as well as out of equilibrium $c$ interacting
with $d$) fail to generate sufficient baryogenesis for similar
reasons.

If we do not assume thermal equilibrium initial conditions (before
relevant species go out of equilibrium), then as before we may
consider the $b$ being in equilibrium while the $c$ quark is out of 
equilibrium. According to our explanation of the enhancement for this 
scenario before, the larger the mass
of the particle that stays in equilibrium until temperature drops to
around $0.2$ GeV, the smaller the exponentially suppressed density of
nonrelativistic particles, and then as long as the self-annihilation
determines the out of equilibrium time $t_c$, the longer is the time
$t_c$ over which baryon number is generated, leading to a larger {\em
accumulation}\/ of baryon asymmetry.  Since $m_b>m_c$ in this scenario,
we thus expect larger accumulation of baryon asymmetry over the
entropy in this scenario.  In fact, the self-annihilation reaction
becomes small by the temperature $T=0.2$ GeV to the extent that it no
longer controls the out of equilibrium time (instead it is determined
by the $b$-decay).
The competing entropy generation is suppressed because of $|V_{cb}|^2
\sim 10^{-3}$ for the weak interactions, and other reactions
generating entropy are insufficient as well in diluting the baryon
asymmetry.

\begin{figure}
{\centering \includegraphics{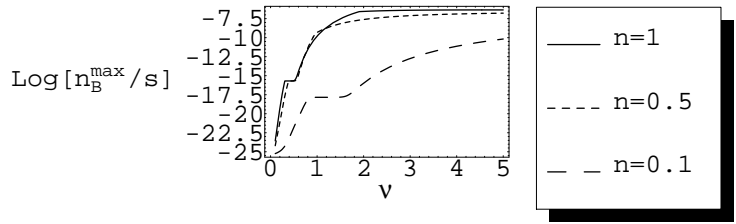} \par}
\caption{\label{fig:enough} 
A conservative upper bound for the baryon
asymmetry generated by out of equilibrium $c$ quarks interacting with
$b$ quarks. The parameters $n$ and $\nu$ are as before. For nonstandard 
cosmologies, sufficient baryogenesis may be possible.  However, as 
explained in the text, the out-of-equilibrium condition seems 
difficult to maintain.}
\end{figure}

For completeness, one should note that since $m_f < m_{f_2}$ in this
scenario there is a contribution to the entropy not accounted for by
\eqr{eq:mainnumeq} and \eqr{eq:alternate}.  This comes from the fact
that $b$ undergoes out-of-equilibrium decay to $c$ even though $b$ has
a chemical equilibrium distribution.  That is because $c$ by
definition is out of equilibrium and thus cannot give the inverse
decay reaction necessary for the $b$ decay to conserve entropy.
However, we can check explicitly this contribution to the entropy 
contribution is negligible, even though it is proportional to the
integration time $t_c$, because of the smallness of the decay rate.
Also, since the $c$ is out of equilibrium, the reaction of $b$-$\bar{b}$
annihilation into $c$-$\bar{c}$ will also contribute to the creation
of new entropy. However, one can check that his contribution is also
insufficient to dilute away the baryon asymmetry generated. The
numerical bounds shown for this scenario in Fig.~\ref{fig:enough}
indicate that this scenario is viable at the level of our analysis.

The most obvious difficulty with this scenario is that another
mechanism is required to set the initial density of $b$ quarks to be 
much larger than that of $c$, as well as to maintain that
hierarchy.  For example, for temperatures far above $m_c$, the gluons
which keep $b$ quarks in equilibrium will tend to produce a large
number of $c$ quarks that will be close to chemical equilibrium.
Since a high initial temperature is crucial, some mechanism besides
just the expansion of the universe is required to keep $c$ out of
equilibrium while is $b$ in equilibrium.  Even if the $c$ densities could
somehow be kept small, since the self-annihilation cross sections for 
$b$ and $c$ are approximately the same, if the cosmological expansion 
is responsible for the out-of-equilibrium condition, then it must 
delicately tune itself to be smaller than the $b$ scattering rate while 
larger than that of the $c$'s.  However, our ``model-independent'' 
analysis does not rule out this scenario, and it may be viable in a 
complicated and fortuitous setting.

As for the scenario in which $b$ quarks are in equilibrium while $u$
quarks are out of equilibrium, it will be even more difficult (if not
impossible) to keep $u$ quarks out of equilibrium for the required
``exponentially long'' duration $t_c$, because the gluons which are
supposed to be in equilibrium have one of their strongest
equilibration reactions through $u$-$\bar{u}$ creation \footnote{It
is trivial to show that particle $X$ cannot have approximately
equilibrium distribution if the particles with which $X$ interacts
{\em most strongly} are not in equilibrium.}. For the scenario with
other quarks (besides the top and the bottom) in equilibrium during
baryogenesis, the exponential enhancement of the duration of
baryogenesis is too weak to allow sufficient baryogenesis.

Hence, remarkably, only one scenario, namely out of equilibrium $c$
quarks interacting with equilibrium $b$ quarks, may generate
sufficient baryon asymmetry with the dimension 6 operator if the out
of equilibrium mechanism can be appropriately engineered.  However, it
would be surprising if such a realistic scenario can be constructed.

\section{Conclusion}
We have studied a new scenario of baryogenesis involving dimension-6
baryon number-violating operators that would generically arise from
integrating out ultraviolet degrees of freedom in the context of the
Standard Model. The question to be answered in this scenario was
whether a time-dependent dimensionless suppression coefficient for the
dimension-6 operator can allow sufficient baryogenesis between the
effective temperatures of $30\,$GeV and $0.2\,$GeV, if the only fields
that are allowed to participate in the short distance physics are
those of the Standard Model. The upper bound on the temperature is
motivated from the cosmological bound on KK graviton decay while the
lower bound is purely for our calculational feasibility (related to
the QCD phase transition). The time-dependent suppression coefficient
was motivated from the generic possibility that with the
Arkani-Hamed-Schmaltz mechanism of suppressing proton decay, the
proton decay suppression factor can be time-dependent.  

Unfortunately,
if the fundamental scale setting the size of the unsuppressed
dimension-6 operator is forced up to values $M_f>100\,$TeV by the bound
from neutron-antineutron oscillations, the branching ratio of
$B$-violating interactions to $B$-conserving interactions is set by
the quantity $(180\,{rm GeV}/M_f)^4 \leq 10^{-13}$.  We have shown
this to be too small to allow sufficient baryogenesis for most cases.
Another way to view this bound is that as the dimension-6 operator
creates baryon asymmetry, the same reactants participate in creating
entropy through $B$-conserving operators. The competition between the
two types of reactions is almost always dominated by the
$B$-conserving one.  Hence, within the perturbative setting where the
4D effective action is valid, sufficient baryogenesis in this scenario
is in most cases impossible almost independently of cosmology.

There is one scenario which cannot simply be ruled out by the general
analysis.  This involves out of equilibrium $c$ quarks interacting
with equilibrium $b$ quarks from a temperature of about $30\,$GeV until
a temperature of $0.2\,$GeV. The reason why this scenario evades the
suppression is that the primary entropy generation channel is through
the $b$-$\bar{b}$ self-annihilation into $c$-$\bar{c}$ instead of the
$b$-$c$ weak interaction, since $|V_{cb}|^2 \sim 10^{-3}$.
Furthermore, given that nonstandard cosmology of braneworlds may give
a slow expansion rate, one can have baryogenesis persist for a long time
period, giving a large integrated pileup.  However, given that $c$
quarks are produced efficiently both by the gluons that keep the $b$
quarks in equilibrium as well as the $b$ quarks which undergo decay,
setting up the necessary hierarchy in densities to keep $c$ out of
equilibrium while $b$ in equilibrium seems difficult.  However, this
scenario does pass the general tests not relying upon the specifics of
the out of equilibrium mechanism.

Since neutrinos are almost massless and have vanishingly small tree
level annihilation cross section in the limit of small $\sqrt{s}$, one
would na\"\i vely think that this channel might be weak enough for the 
baryon number-violating operator to compete with it. Unfortunately, this
turns out not to be the case. 

It is interesting to note that even with just the Standard Model
fields, the presence of nonrenormalizable operators implies the existence 
of additional $CP$-violating rephasing invariants beyond the Jarlskog 
parameter. In any theory, the number of 
physical $CP$-violating phases is given by the number of couplings in 
the Lagrangian that are allowed to be complex, minus the dimension of the 
symmetry group which describes phase redefinitions of the fields: 
introducing new effective operators into the SM automatically creates more 
$CP$-violating invariants, even if all the couplings of the new operators 
are real in some basis. Hence, it is not the smallness of the
Jarlskog invariant that prevents the success of our baryogenesis
scenario. It is merely the fact that unless neutron-antineutron
oscillations are suppressed by some means other than the fundamental 
scale, the phenomenologically acceptable scale of $100\,$TeV
is still too large for dimension-6 operators to play a significant
role for baryogenesis below the temperature of the electroweak phase 
transition. 

As with any ``no go'' arguments, there are many loopholes in our
conclusion (besides the one remotely possible scenario that we already
mentioned).  First, we neglected any possible nonperturbative physics.
The 4D coupling constants are inversely proportional to the volume of
the extra dimensions in which the gauge fields propagate. Thus, if the
volume of the extra dimensions was small at some epoch in the early
Universe, then nonperturbative physics may dominate (see Appendix
\ref{app:inst}), given that the perturbative contributions to $n_B/s$
are insufficient in our scenario. For example, the $SU(2)_L$ instanton
effects may become unsuppressed, giving rise to a scenario similar to
that of sphaleron transitions at the electroweak phase transition.
Also, for effective temperatures below about $0.2$ GeV,
non-perturbative QCD effects will be relevant, since the quark degrees
of freedom become confined. Although this is not likely to change the
suppression of the dimension-6 operator, the kinematics and matrix
elements will be very different from our perturbative calculation done
in the regime of deconfined quarks.

Secondly, we have neglected all effects of the KK modes; or, in the 
higher dimensional picture, we have neglected the fact that the space
may be inhomogeneous in the higher dimensions.  For example, one may
envisage a scenario in which there is a phase transition which
localizes the wave functions of the quarks and the leptons.  A proper
description of quark-lepton separation in higher dimension will require
analysis beyond the zero mode.  However, this type of effect will
most likely play a role in enhancing the nonequilibrium condition rather
than changing the baryon number-violating branching ratio.  Hence, our
conclusion is most likely robust with respect to this assumption.

Thirdly, we have neglected all CPT violation effects that must exist
because of the time dependence of the quark-lepton separation.  If the
time scale associated with the quark-lepton separation process is very
short, then there may be significant contributions from these effects
to enhance the baryon number violating channels.  However, such models
will probably be severely constrained by the restriction on bulk
graviton production, just as the reheating temperature is severely
constrained.

Hence, although there are loopholes and caveats to our ``no go''
claims, it seems fair to conclude that with only perturbative physics,
degrees of freedom beyond the Standard Model fields must play a 
significant role in baryogenesis if the effective temperature of the 
Universe never exceeds 30 GeV (which is a very conservative upper bound
for the maximum temperature in models with large extra dimensions, if
the fundamental scale is to be accessible to collider experiments). 
The requirement of beyond the Standard Model field content
is not an obvious statement, considering that we hardly impose any
restrictions on the cosmology and that there is an unsuppressed
dimension-6 operator as well as an effectively large $CP$ violating
phase available for baryon asymmetry generation.  Of course, our ``no
go'' claims would be significantly relaxed if the dimension-9 operator
responsible for neutron-antineutron oscillations is suppressed by some
symmetry mechanism which is introduced by hand to supplement the 
geometrical suppression of proton decay operators.

In closing, note that even if the fundamental scale
can be lowered by evading the bound from neutron-antineutron oscillation,
there is a significant challenge in this type of scenario as the
quarks (and/or leptons) must be forced out of equilibrium at rather high
temperatures. To accomplish this, it is possible to lower the Planck
scale during baryogenesis by having a small extra
dimensional volume during that period.  However, there are severe
restrictions coming from bulk graviton production in such cases.
Furthermore, it is an extremely difficult challenge to find a time
dependent potential for the scalar fields localizing the quarks and
the leptons, such that the initial quark-lepton separation departs
sufficiently from today's equilibrium value in a natural manner.  We
leave variations on our model that may generate sufficient baryon
asymmetry to future studies.

\acknowledgments
We thank Aaron Grant, Nima Arkani-Hamed, Misha Shaposhnikov, Lisa Everett, 
Rocky Kolb, Antonio Riotto, Marty Einhorn, Gordy Kane, Antonio Masiero, Tony
Gherghetta, Alexander Dolgov, Karim Benakli, Gianfrancesco Giudice and 
Rula Tabbash for helpful conversations and Boris K{\" o}rs for 
correspondence.
This research was supported in part by DOE Grant DE-FG02-95ER40899, Task G.

\appendix
\section{Adequate suppression of proton decay} \label{app:pprotection}
To find the minimum value of $L$ consistent with proton stability, we consider the decay process in the four-dimensional low energy theory, where we have argued that four-fermion operators give the largest contribution. (Processes that cannot be written as four-fermion effective operators, for example wormholes, are negligible for the value of $L$ that we find is needed.) The proton decay rate calculation for these operators is analogous to muon decay in the Fermi theory, giving an estimate of\begin{equation}
\tau_p \approx (\lambda v_s/M_f^2)^{-2} m_p^{-5} = (\lambda v_s)^{-2} a^2 M_f^4 m_p^{-5} \label{tau_p}
\end{equation}
for the ``partial lifetime'' of decay into any allowed channel, where $\lambda$ is a dimensionless constant of order 1. Mode-dependent experimental lower limits on the proton lifetime \cite{Groom:in} require $\tau_p>2\times 10^{31}$ years for some modes, which would translate into \mbox{$M_f>3\times 10^{15}$ GeV} in the absence of the suppression factor $v_s$. Thus 
\begin{equation}
a \leq (M_f\cdot 3\times 10^{15}\,\rm{GeV})^2 \approx 10^{21-23} \approx e^{48-53}
\end{equation} 
where the range quoted is for a fundamental scale ranging from 100 to $10\,$TeV. The conclusion is that about 50 e-foldings of suppression are required at the present epoch. This is easy to obtain if $v_s\sim e^{-\mu^2 L^2}$ and $\mu$ is not too small relative to $M_f$, but is marginally consistent with perturbativity of SU$(2)_{W}$, which begins to break down at $M_f R_5>30$ (recall that $R_5 \geq L$), for proton decay mediated by states of mass $M_f$, for which $a\sim e^{-k M_f L}$ with $k$ of order 1.

\section{Other potential sources of $B$ violation} \label{app:inst} 
One should also consider nonperturbative processes which may affect
$B$ violation and baryogenesis \cite{KuzminRS,Rubakov:1996vz}. Even 
without the introduction of $B$-violating operators, the baryon number 
current is only conserved up to a SU$(2)_W$ anomaly, and it is well-known 
that topologically nontrivial gauge configurations can lead to a change 
in $B+L$, either by quantum tunneling via instantons, or thermal
excitation via sphalerons \footnote{The mixed U$(1)_B\times{\rm
U}(1)_Y^2$ anomaly is also nonvanishing, but there are no nontrivial
U$(1)$ gauge configurations analogous to the instanton or
sphaleron.}. The rates for $B$ violation by instantons and sphalerons
also have exponential suppression factors $e^{-8\pi/g_2^2}$ and
$e^{-M_{\rm S}/T}$ respectively, where $M_S$ is the sphaleron energy
given by $M_{\rm S}\approx 2M_W/\alpha_W$ \footnote{Hill and Ramond
have reinterpreted the four-dimensional instanton rate as the
amplitude for a massive soliton in five-dimensional gauge theory to
propagate over the entire fifth dimension \cite{HillRamond}.}. If the
total length of the fifth dimension $R_5$ were to change with time
then, assuming a constant five-dimensional gauge coupling $g_5$, these
processes could become important. Given the relationship
$g_4^2=g_5^2/R_5$ for the weak gauge coupling, instantons would be
exponentially larger for smaller $R_5$; the dependence is just
$e^{-{\rm const.}\times R_5}$, so the instanton enhancement at small
$R_5$ would shadow the increase in processes mediated by massive
states. However, it can easily be checked that the instanton rate
today is many orders of magnitude smaller than the bound on such
perturbative processes, therefore instantons can never have dominated
over the perturbative operators, unless the perturbative operators
were further suppressed by many orders of magnitude (for example, if
they vanished by an exact discrete symmetry). The sphaleron rate is
more difficult to estimate, since in addition to the temperature
dependence one would have to model how the Higgs v.e.v.\ changes in
response to changing $R_5$. On the simplest assumption of a constant
v.e.v.\ the sphaleron rate is actually independent of $R_5$ and, as in
four dimensions, depends just on $T$. This is a counterexample to the 
claim that all processes are exponentially suppressed with increasing 
$R$; it is understood by noting that the formation of sphaleron 
configurations is a classical, thermally-activated process, in contrast 
to quantum effects which are suppressed by a massive propagator or by 
the Euclidean action. Unsuppressed sphaleron transitions ``wash out'' 
any baryon asymmetry generated at temperatures above the electroweak 
scale \footnote{Unless a lepton asymmetry associated with the conserved 
$B-L$ number is generated at high scales.}; on the assumption of a very 
low maximum temperature of the Universe, which is required in low-scale 
models, sphaleron effects will be completely negligible.

\section{Boltzmann equations and entropy
\label{appendixboltzmann}}
Consider first the entropy of one species of particles. Divide the
phase space of particles into cells and let the cells be numbered \(
i=1,2,3,... \) where each cell contains \( N_{i} \) identical
particles. Let \( G_{i} \) be the phase space volume that can be occupied
by one particle in the cell \( i \). The total phase space of the \( i
\)'th cell is
\begin{equation}
\Delta \Gamma _{i}=\frac{G_{i}^{N_{i}}}{N_{i}!}\end{equation}
if one assumes \( 1\ll N_{i}\ll G_{i} \) and negligibly correlated 
distributions.
The total phase space volume is 
\begin{equation}
\Delta \Gamma =\prod _{i}\Delta \Gamma _{i}\end{equation}
and the definition of entropy is \( S=\ln \Delta \Gamma  \). Using the approximation
\( \ln N!\approx N\ln (\frac{N}{e}) \), we can express the entropy as
\begin{eqnarray*}
S & = & \sum _{i}G_{i}\frac{N_{i}}{G_{i}}\ln [\frac{eG_{i}}{N_{i}}]\\
 & \approx  & \int \frac{d^{3}pd^{3}x}{(2\pi )^{3}}f(p,x)\ln \left[\frac{e}{f(p,x)}\right]
\end{eqnarray*}
where
\begin{equation}
n(x)\equiv \int \frac{d^{3}p}{(2\pi )^{3}}f(p,x)\end{equation}
gives the number density of particles. Hence, one can define the
entropy current
\begin{equation}
s^{\mu }\equiv \int \frac{d^{3}p}{(2\pi )^{3}}f(p,x)\ln \left[\frac{e}{f(p,x)}\right]\frac{p^{\mu }}{p^{0}}
\end{equation}
where we define the entropy density as usual as 
\begin{equation}
s\equiv s^{0}.
\end{equation}
Now, using the metric
\begin{equation}
ds^{2}=dt^{2}-a^{2}(t)d\vec{x}^{2}\end{equation}
one has 
\begin{equation}
s^{\mu }_{\, \, \, ;\mu }=\dot{s}+3Hs\end{equation}
 and hence the entropy density Boltzmann equation becomes
\begin{equation}
\label{eq:entropyboltzmann1}
\dot{s}+3Hs=-\sum _{f}\sum _{i}\int d\,\Gamma C_{i}^{f}[f_{f}]\ln [f_{f}]\end{equation}
where \( C_{i}^{f} \) corresponds to \( i \)th collision operator for the
particle species \( f \) in the Boltzmann equation and \( d\,\Gamma  \) represents
the appropriate momentum phase space integral. Neglecting \( CP \) violations
which are not important for the total entropy generation, we can write the collision
terms in terms of S-matrix amplitudes as 
\begin{eqnarray}
-\sum _{f}\sum _{i}\int d\,\Gamma C_{i}^{f}[f_{f}]\ln [f_{f}] & = & \sum _{i}\int d\,\Gamma (2\pi )^{4}\delta ^{(4)}(\textrm{momenta})\times \nonumber \\
 &  & (f_{a_{i}}f_{b_{i}}-f_{c_{i}}f_{d_{i}})|M(c_{i},d_{i};a_{i},b_{i})|^{2}\ln [\frac{f_{a_{i}}f_{b_{i}}}{f_{c_{i}}f_{d_{i}}}]+\nonumber \\
 &  & \sum _{i}\int d\,\Gamma (2\pi )^{4}\delta ^{(4)}(\textrm{momenta})\times \nonumber \\
 &  & (f_{e_{i}}-f_{g_{i}}f_{h_{i}})|M(g_{i},h_{i};e_{i})|^{2}\ln [\frac{f_{e_{i}}}{f_{g_{i}}f_{h_{i}}}]+\nonumber \\
 &  & \sum _{i}\int d\,\Gamma (2\pi )^{4}\delta ^{(4)}(\textrm{momenta})\times \nonumber \\
 &  & (\textrm{initial states}-\textrm{final states})|M|^{2}\ln [\frac{\textrm{initial states}}{\textrm{final states }}]+...\nonumber \\
 &  & +\Delta C\label{eq:entropyboltzmann} 
\end{eqnarray}
 where the phase space integral 
\begin{equation}
d\,\Gamma \equiv \prod \frac{g}{(2\pi
 )^{3}}\frac{d^{3}p}{2E}\end{equation} is over all particles and
 momenta, and \( \Delta C \) corresponds to the entropy change
 produced by the decay of a classical field configuration. If a
 zero-temperature classical field configuration (such as an inflaton
 vev) decays to particles \( X_{1} \) and \( X_{2} \) and the
 Bogoliubov coefficient for the particle production is \( |\beta
 _{k}|^{2} \) (where the physical momentum is \( k/a \)), then we have
 the effective collision operator
\begin{equation}
\Delta C=\int d\,\Gamma C_{i}^{X_{1}}[f_{X_{1}}]\ln [f_{X_{1}}]+\int d\,\Gamma C_{i}^{X_{2}}[f_{X_{2}}]\ln [f_{X_{2}}]=-\int \frac{d^{3}p}{(2\pi )^{3}}\ln [f_{X_{1}}f_{X_{2}}]\frac{d}{dt}|\beta _{pa}|^{2}
\end{equation}
contributing to the entropy density Boltzmann equation.

With our expression for the entropy, one can prove the following
little theorem:  
If baryons are generated through particle \( f \) which is out of equilibrium
undergoing arbitrary reactions, then 
\begin{equation}
n_{B}\leq \frac{2\eta }{a^{3}}\int dt\,\Gamma _{\rm BV}(t)a^{3}n_{f},
\label{eq:rigorous}
\end{equation}
and
\begin{equation}
\label{eq:roughbound}
\frac{n_{B}}{s}<2\eta \left[ \frac{\Gamma _{\rm BV}(t)}{\Gamma _{\textrm{t }}(t)}\right] _{\rm max}
\end{equation}
where 
\begin{equation}
\eta \equiv \alpha _{W}\sum _{ij} I \,\textrm{Im}\,\{\lambda
^{ajbi}\lambda ^{*atbs}V_{is}V^{*}_{tj}\}\sim 0.02 \frac{\bar{T}^2}{
v^2},
\end{equation} 
$\bar{T}$ is the effective energy scale which characterizes the
kinetic energy of the on shell particles in the loop \footnote{More
specifically, $\bar{T}^2$ is approximately $T_{init}^2 + m_{init}^2 -
(m_{ui} + m_{dj})^2$ where $T_{init}$ and $m_{init}$ are temperature
and mass scales of the initial reactants and $m_{ui}$ and $m_{dj}$ are
the masses of the particles in the loop.}, $I$ is the loop integral
suppression factor, $v$ is the Higgs VEV of about $180$ GeV, and \(
\Gamma _{\rm BV} \) is the tree-level baryon number violating reaction
rate.  The maximum in \eqr{eq:roughbound} is taken during when most of
the baryogenesis is taking place, and \( \Gamma _{\rm t} \)
corresponds to the reaction rate of \( f \) which does not violate
baryon number.

The argument validating this claim goes as follows.  Starting from
Eq.~(\ref{eq:genericbaryonevolve}), one can use the momentum
conserving $\delta$-function and the form of the equilibrium
distribution to obtain \eqr{eq:rigorous} if one assumes that the 
baryon number-violating operator coupling is of order 1 (maximal).
Now, consider the \( f \) evolution. Defining \( N_{f}\equiv n_{f}a^{3} \),
we can write the $B$-conserving part of the Boltzmann equation as 
\begin{equation}
\dot{N}_{f}=-\sum _{i}\Gamma ^{i}(N_{f}-N_{f}^{eq}U^{i}_{f})+\Delta Ca^{3}+...\end{equation}
where \( \Gamma ^{i} \) are baryon number-conserving reaction rates, 
\( N_{f}^{eq} \)
are equilibrium number densities per comoving volume, \( U^{i}_{f} \) are the 
ratios of final state number densities to initial state equilibrium number 
densities that usually appear in the Boltzmann equations, and 
\( \Delta Ca^{3} \) is the external source for \( f \). Since we are assuming 
maximal out of equilibrium, the equilibrium terms involving reverse reactions 
can be neglected. Hence, we can write
\begin{equation}
\dot{N}_{f}\approx -\Gamma _{\rm t}(t)N_{f}+\Delta Ca^{3}\end{equation}
 which governs \( N_{f} \) neglecting the baryon number violating terms. Hence,
we can write
\begin{equation}
n_{B}\leq \frac{2\eta }{a^{3}}\int _{t_{i}}^{t}dt'\frac{\Gamma _{\rm BV}(t')}{\Gamma _{\rm t}(t')}(\Delta C(t')a^{3}(t')-\dot{N}_{f}(t'))\end{equation}
Furthermore, we  can write
\begin{equation}
n_{B}\leq \frac{2\eta }{a^{3}}\left[\frac{\Gamma _{\rm BV}}{\Gamma
_{\rm t}}\right]_{\rm max} \left(\int _{t_{i}}^{t}dt'\,\Delta
C(t')a^{3}(t')-N_{f}(t)+N_{f}(t_{i})\right) \end{equation}
From Eq. (\ref{eq:entropyboltzmann1}), we see that
\begin{equation}
sa^{3}\geq N_{f}(t_{i})+\int _{t_{i}}^{t}dt'\Delta C(t')a^{3}(t')\end{equation}
Hence, we find Eq.~(\ref{eq:roughbound}), as promised.

\section{Scaling of temperature}
\label{appendixscaling}
Here we discuss the physical range of \( \nu  \) in \( T\propto a^{-\nu } \).
The generalized first law of thermodynamics tells us
\begin{equation}
dE=-PdV+TdS+J_{\rho }dt
\end{equation}
where \( J_{\rho } \) is the external source/sink of energy for the photons
and for other relativistic fluids in thermal equilibrium with the photons. Here we
have separated the 3 dimensional pressure \( P \) from the component associated
with the extra dimension, whose contribution to \( dE \) we write as 
\( J_{\rho}\,dt \).
If we assume a constant ``effective'' average equation of state 
\( P/\rho \approx \omega \), set \( dS=0 \), and assume that \( dV=a^{3} \), 
then we find, trivially,
\begin{equation}
\frac{\rho }{\rho _{i}} = \exp\left(\int \frac{J_{\rho }}{a^{3}\rho }dt\right)
\left(\frac{a}{a_{i}}\right)^{-3(\omega +1)}.
\end{equation}
For photons and other relativistic particles in thermal equilibrium confined
to 3 spatial dimensions, the equation of state is \( \omega \approx 1/3 \)
and \( \rho \propto T^{4} \). This implies
\begin{equation}
T = T_{i} \exp\left(\frac{1}{4}\int \frac{J_{\rho }}{a^{3}\rho }dt\right) \left(\frac{a}{a_{i}}\right)^{-1}
\end{equation}
which implies that \( \nu =1 \) unless the source term 
\( \int dt\,\frac{J_{\rho }}{a^{3}\rho } \) is significant. 
If \( J_{\rho }>0 \), then \( \nu <1 \), otherwise \( \nu >1 \).

An example of a situation in which \( \nu <1 \) is when the bulk fields are
in thermal equilibrium with the brane fields through
\begin{equation}
J_{\rho }dt\approx -P_{4+d}\,a^{3}dz^{d}\end{equation}
where \( P_{5} \) is the bulk pressure and \( z \) is the scale factor for
the extra dimension.

\end{document}